# Plasma-engineered Hydroxyl Defects in NiO: a DFT-Supported-Spectroscopic Analysis of Oxygen-Hole States and Implications for Water Oxidation

Harol Moreno Fernández[1*], Mohammad Amirabbasi[2*], Crizaldo Jr. Mempin[1,3], Andrea Trapletti[4], Garlef Wartner[5], Marc F. Tesh[5], Esmaeil Adabifiroozjaei[6], Thokozile A. Kathyola[7], Carlo Castellano[4], Leopoldo Molina-Luna[6], and Jan P. Hofmann[1*]

1) Surface Science Laboratory, Department of Materials and Geosciences, Technical University of Darmstadt, Peter-Grünberg-Straße 4, 64287 Darmstadt, Germany
2) Institute of Materials Science, Materials Modeling, Technical University of Darmstadt, Otto-Berndt-Straße 3, 64283 Darmstadt, Germany
3) Electrochemical Materials and Interfaces, Dutch Institute for Fundamental Energy Research, 5612 AJ Eindhoven, Netherlands
4) Dipartimento di Chimica, Università degli studi di Milano, Via Golgi 19, 20133 Milan, Italy
5) Max Planck Institute for Chemical Energy Conversion, Stiftstraße 34-36, 45470 Mülheim an der Ruhr, Germany
6) Advanced Electron Microscopy Division, Institute of Materials Science, Department of Materials and Geosciences, Technical University of Darmstadt, Peter-Grünberg-Straße 2, 64287 Darmstadt, Germany
7) Diamond Light Source, Didcot, Oxfordshire, OX11 0DE, United Kingdom

**Corresponding authors**

Harol Moreno Fernández – hmoreno@surface.tu-darmstadt.de
Mohammad Amirabbasi – amirabbasi@mm.tu-darmstadt.de
Jan P. Hofmann – hofmann@surface.tu-darmstadt.de

# Abstract

Controlling lattice-oxygen reactivity in earth-abundant OER catalysts requires precise tuning of defect chemistry in the oxide lattice. Here, we combine DFT+$U$ calculations with plasma-assisted synthesis to show how $O_2$ and $H_2O$ in the discharge govern vacancy formation, electronic structure, and catalytic predisposition in NiO thin films. Oxygen-rich plasmas generate isolated and clustered Ni vacancies that stabilize oxygen-ligand-hole states and produce shallow O 2$p$–Ni 3$d$ hybrid levels, enhancing Ni–O covalency. In contrast, introducing $H_2O$ during growth drives local hydroxylation that compensates vacancy-induced $Ni^{3+}$ centers, restoring $Ni^{2+}$-like coordination, suppressing deep divacancy-derived in-gap states, and introducing shallow Ni–O–H–derived valence-band tails. EXAFS confirms that hydroxylation perturbs only the local environment while preserving the medium-range NiO lattice, and Ni L-edge spectroscopy shows a persistent but redistributed ligand-hole population. These complementary vacancy- and hydroxylation-driven pathways provide a plasma-controlled route to pre-define electronic defect landscapes in NiO and to tune its activation toward OER-relevant NiOOH formation.

## 1. Introduction

Hydrogen and oxygen production via water electrolysis (i.e., water splitting) requires catalysts that are both active and stable, as well as electrically conductive under operating conditions. Precious metal-based catalysts, such as platinum and iridium dioxide, are effective for catalyzing water splitting in acidic media; however, they are expensive and scarce[1]. In contrast, the oxygen evolution reaction (OER) in alkaline environments is typically performed using non-noble transition-metal oxide (TMO) catalysts[2]. However, these generally exhibit lower activity compared to their noble-metal counterparts in acidic media. As a result, significant efforts have been directed toward developing inexpensive, efficient, earth-abundant, and environmentally benign electrocatalysts for these reactions[3]. Nickel oxide (NiO) is a promising electrode material for the alkaline OER due to its low cost, natural abundance, and favorable redox properties[4]. As a p-type semiconductor with a wide energy gap (>3.0 eV)[5], NiO features partially filled Ni $3d$ states that hybridize with O $2p$ orbitals, providing a basis for redox activity. When immersed in an alkaline electrolyte, its surface readily converts to nickel hydroxide ($Ni(OH)_2$), an n-type semiconductor with a wide energy gap (>3eV)[6,7], which, under anodic potentials, further transforms into the OER-active nickel oxyhydroxide (NiOOH, p-type semiconductor with a reduced energy gap)[6,8].

Ni hydroxide/oxyhydroxide films are generally accessible via a redox reaction that involves electrochemical activation of NiO films[9,10] –which are typically prepared by dry, vacuum-based processes such as sputtering and pulsed laser deposition (PLD)– in an aqueous electrolyte solution. Nevertheless, only a few reports describe the direct preparation of NiOOH and $Ni(OH)_2$ films using dry methods[11,12]. Separately, several wet-chemical routes–such as sol-gel, dip-coating, and cathodic deposition–can also yield Ni hydroxides and oxyhydroxides[13–15]. Nonetheless, these reports often lack correlative spectroscopic evidence or systematic comparisons to confirm phase formation as-deposited. This absence of data leaves uncertainty as to whether hydroxide or oxyhydroxide phases, and the associated oxygen-hole states, can be stabilized before electrochemical treatments. Besides, even if dry or wet processes successfully form the NiOOH phase, its instability in air presents a significant challenge for its characterization; even brief exposure to the atmosphere during transfer can degrade NiOOH; thus, by the time the sample is returned to vacuum to be analyzed spectroscopically, the surface may no longer reflect its initial state, as we have shown in recent studies[16,17]. Therefore, unambiguous identification of the active phase requires vacuum-integrated approaches such as X-ray photoelectron spectroscopy (XPS), ensuring that the deposited phase can be analyzed without ambient degradation.

A further challenge lies in the electrochemical conditioning typically required to activate NiO. Converting NiO into $Ni(OH)_2$ often demands tens to hundreds of cyclic voltammetry (CV) cycles before stable properties are achieved[16,18]. This activation period is critical, as larger amounts of $Ni(OH)_2$ facilitate stabilization of the catalytically active NiOOH phase[10,17]. Therefore, introducing hydroxyl groups during deposition– for example, by incorporating $H_2O$ vapor during growth– not only accelerates this transformation but also promotes the formation of mixed Ni–O–H coordination environments, as we will show in later sections.

$Ni^{3+}$–$O^-$ configurations reflect the broader role of lattice oxygen redox, in which oxygen-hole states ($O^-$ species) emerge from strong Ni–O covalency and directly participate in O–O bond formation[19,20]. Nevertheless, stabilizing such oxygen holes remains a central challenge: they typically form only under electrochemical bias, are dynamically coupled to structural changes, and are difficult to embed during synthesis without compromising stability[21,22]. Under oxygen-rich growth conditions, Ni vacancies act as the dominant charge-compensation mechanism and can stabilize $Ni^{3+}$–$O^-$ configurations by introducing mixed O $2p$–Ni $3d$ states near the Fermi level. In contrast, when hydroxyl species are incorporated (e.g., under $H_2O$-containing plasma conditions), $OH^-$ ligands tend to compensate vacancy-induced $Ni^{3+}$ centers and restore local $Ni^{2+}$ coordination, thereby suppressing the formation of additional lattice oxygen hole ($O^-$) states. To rationalize these trends at the electronic-structure level, density-functional theory (DFT) reveals how cation vacancies and incorporated hydroxyl species tune the mixed O $2p$–Ni $3d$ states near the Fermi level in NiO. These calculations predict that introducing Ni vacancies generates in-gap states, narrowing the electronic gap[23], whereas hydroxylation modifies Ni–O covalency by creating shallow Ni–O–H hybrid states rather than stabilizing new ligand-hole configurations. These defect-induced states exhibit a mixed O $2p$–Ni $3d$ character near the Fermi level, enhancing Ni–O covalency, and in oxygen-rich films can stabilize oxygen-hole configurations that lower the barrier for lattice oxygen participation during OER. However, experimental realization and spectroscopic validation of these predicted states remain scarce. Thus, strategies that directly tune Ni–O hybridization–whether through vacancy formation or controlled hydroxylation–provide a promising pathway to engineer lattice-oxygen-mediated OER in TMOs[19,24].

In this study, we integrate plasma-assisted synthesis of NiO with density functional theory (DFT) and experimental spectroscopy to establish how $O_2$ and $H_2O$ in the discharge control the formation of Ni vacancies, ligand-hole states, and hydroxylated coordination environments. Controlled plasma deposition allows systematic tuning of vacancy density and hydroxyl content, which in turn modulate Ni–O covalency and the defect-derived electronic states near the valence band edge. Under oxygen-rich conditions, DFT predicts that vacancy clustering can stabilize $Ni^{3+}$–$O^-$ configurations– although in the experimentally grown 88% $O_2$ films these Ni-centered holes are largely below the detection threshold and the ligand-hole character appears predominantly O-centered–whereas $H_2O$ incorporation compensates vacancy-induced $Ni^{3+}$ and restores local $Ni^{2+}$ coordination without generating additional ligand holes. This behavior modifies the surface electronic structure, facilitating the NiO → $Ni(OH)_2$ → NiOOH transformation during electrochemical conditioning. Consequently, the combined plasma–DFT approach provides a controlled means to pre-define defect configurations during synthesis and to tailor the activation behavior of NiO.

The manuscript is organized as follows. Section 2.1 presents the DFT framework used to evaluate how Ni vacancies and hydroxyl species modify the electronic structure of NiO. Section 2.2 describes the plasma-assisted synthesis of NiO and h-NiO thin films and their structural characterization. Section 2.3 provides the XPS and valence band analysis that resolves the evolution of Ni–O, Ni–O–H. Section 2.4 discusses the Ni K-edge XANES/EXAFS results, highlighting local and medium-range structural changes associated with vacancy formation and hydroxylation. Section 2.5 correlates these defects and electronic signatures with the electrochemical OER performance, including Tafel analysis, and turnover frequency (TOF). The final section summarizes the implications of plasma-driven defect engineering for preconditioning NiO-based OER catalysts.

## 2. Results and Discussion

### 2.1. Magnetic Structure and Defect-Induced Electronic States of NiO and h-NiO

First-principles calculations are performed within density functional theory (DFT) using the projector augmented-wave (PAW) basis set[25] as implemented in the Vienna Ab initio Simulation Package (VASP)[26]. We first examine the magnetic moments and spin-resolved density of states (DOS) of stoichiometric NiO. In the pristine supercell, all Ni sites carry essentially identical local environments. However, because PBEsol alone cannot adequately capture strong electron-electron correlations, particularly in Mott insulators such as NiO, we treat the Ni $3d$ states with the DFT+$U$ [27,28] method, applying an on-site Coulomb interaction $U$=4 eV; full parameters are given in the methodology section. With this approach, all Ni sites carry essentially identical local moments of about $\pm 1.61\mu_B$, arising almost entirely from the $3d$ channel, whereas all O sites are essentially non-magnetic with moments numerically indistinguishable from zero, in good agreement with previous calculations[29]. The Ni moments are arranged in the conventional antiferromagnetic (AFM) pattern: one sublattice is spin-up and the other spin-down, so the contributions from the two sets of Ni ions cancel and the total magnetization of the cell is zero. The corresponding spin-resolved total and projected DOS (Figure 1a) is nearly perfectly symmetric with respect to spin and exhibits a clear insulating gap at the Fermi energy. The valence band (VB) is dominated by Ni-$3d$ states with substantial hybridization to O-$2p$, while the conduction band has mainly Ni-$3d$ character.

To link our DFT models to experiment, we assume that NiO films grown at 1.0–20% O$_2$ content in the plasma discharge develop Ni deficiencies, consistent with the XPS evidence discussed later. We therefore introduce a single Ni vacancy ($V_{Ni}^{''}$) on the spin-up sublattice. When this single vacancy is introduced, the overall AFM order remains intact. Most of the remaining Ni ions retain moments close to $\pm 1.61\mu_B$, and only a few Ni atoms in the spin-down layer show a modest reduction of the moment to -1.50$\mu_B$. In contrast to the pristine case, a subset of O ions in the supercell develops small but finite spin polarizations of order 0.04-0.06$\mu_B$. The combination of slightly modified Ni moments and these induced O moments compensates for the missing spin of the removed Ni ion, so that the total magnetization of the defective supercell remains practically zero. The DOS for this single-vacancy configuration (Figure 1b) closely resembles that of pristine NiO. It shows that a finite band gap at the spin-up channel and the total DOS at the Fermi energy are essentially zero. The main effect of the vacancy is a subtle redistribution of DOS in the spin-down channel near the top of the VB, where the O-2$p$ contribution becomes somewhat enhanced. The Fermi energy is pinned at the valence band maximum (VBM), reflecting the presence of holes introduced by the $V_{Ni}''$. However, the absence of appreciable DOS at $E_F$ shows that the system remains in an antiferromagnetic insulator, possibly with weak p-type character, rather than becoming metallic. Our findings for a single $V_{Ni}''$ are consistent with previous theoretical and experimental studies on Ni-deficient NiO, which identify $V_{Ni}''$ as the dominant native acceptor and link it to hole states with mixed Ni-3$d$/O-2$p$ character near the VBM, while the system remains an antiferromagnetic insulator at realistic vacancy concentrations.

Previous DFT+$U$ and hybrid-functional studies of Ni vacancies, including calculations on NiO (100) surfaces, show that removal of a Ni$^{2+}$ cation introduces two holes with predominant O-2p (ligand-hole) character[23,30]. Hybrid functionals tend to localize these holes more strongly on oxygen atoms adjacent to the vacancy, whereas DFT+$U$ descriptions often yield more delocalized hole distributions extending over the Ni–O network. In our bulk PBEsol+$U$ calculations at the vacancy concentration considered here ($U$ = 4 eV), attempts to enforce fully localized O$^-$ or Ni$^{3+}$ configurations via occupation-matrix control relax back to a delocalized hole state, indicating that the holes introduced by a $V_{Ni}''$ remain distributed over the surrounding Ni–O framework rather than being stabilized as discrete O$^-$ or Ni$^{3+}$ centers.

We next consider a supercell containing two spatially separated Ni vacancies–reflecting the high-vacancy conditions produced when NiO is processed with 88% O$_2$ in the plasma discharge– one on each antiferromagnetic sublattice, as far as possible from each other in a 3×3×3 supercell (7.15 Å). The site-resolved spin moments show that most Ni ions still carry moments close to ± 1.61$\mu_B$, as in the pristine system, but two Ni sites exhibit strongly reduced moments of about ± 0.85$\mu_B$. These reduced moments are consistent with a change of the local configuration from high-spin Ni$^{2+}$ (S = 1) to low-spin Ni$^{3+}$. As in the single-vacancy case, some O ions acquire finite moments up to 0.07 $\mu_B$, whereas others stay only weakly polarized. The spin-resolved DOS for the double-vacancy configuration (Figure 1c) denotes additional Ni-3$d$ appearing near the valence band edge, which we associate with Ni$^{3+}$ related states created to compensate for the removal of two Ni$^{2+}$ cations. The Fermi energy is again pinned at the top of the VB. Because the two low-spin Ni$^{3+}$ centers reside on opposite spin sublattices, their moments cancel each other, and the total magnetization of the supercell stays essentially zero, so that the global antiferromagnetic character of NiO is preserved even in the presence of two Ni vacancies. For multiple Ni vacancies, our results are consistent with previous work on Ni-deficient NiO, which shows that increasing the Ni-vacancy concentration mainly enhances $Ni^{3+}$/hole character and adds Ni-3$d$/O-2$p$ spectral weight near the VBM. At the same time, the material remains antiferromagnetic and insulating over a broad range of nonstoichiometries[23,31–33]. These studies also find only moderate modifications of the local Ni moments and no robust metallic phase, in agreement with our double-vacancy configuration, where the global AFM order is preserved.

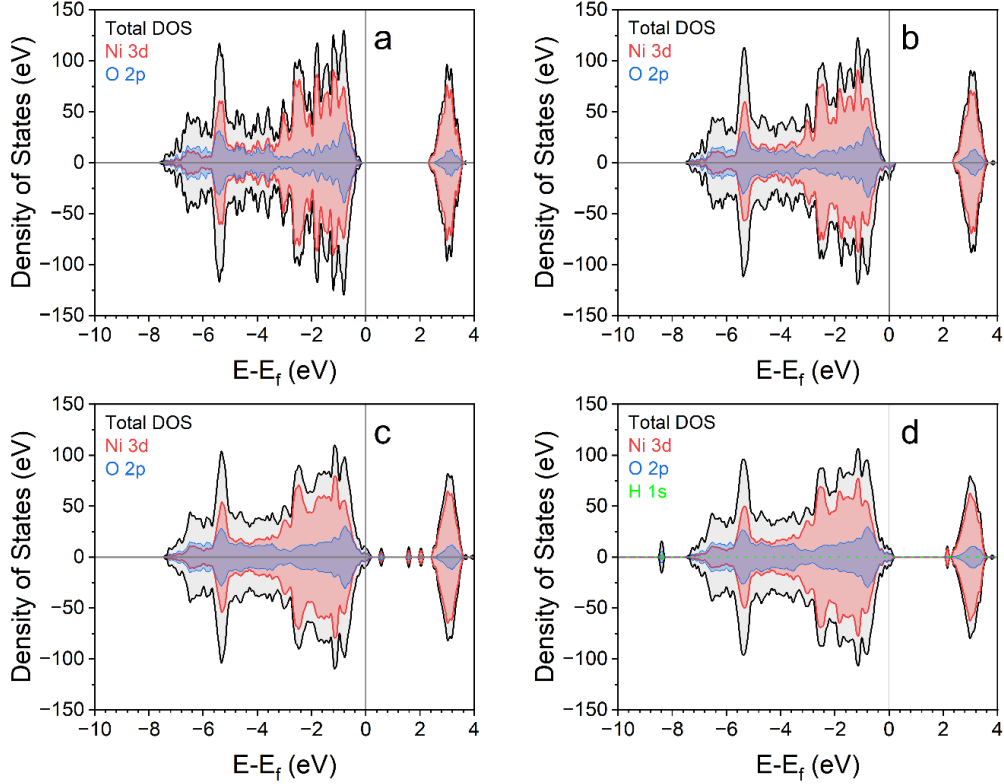

Figure 1 Density of states (DOS) for (a) pure NiO bulk, (b) with one Ni vacancy, (c) with two vacancies, and (d) sample (c) with H-terminations. For two Ni vacancies, one Ni spin up, and one Ni spin down have been removed so that they are in far-field distance. The two nearest neighbors close to Ni-vacancies are oxidized from $Ni^{2+}$ to $Ni^{3+}$. The local spin moments of reduced Ni ions are $+0.81\mu_B$ and $-0.81\mu_B$. Hence, the system is still AFM, and the DOS for spin-up and spin-down channels are symmetric, and the total spin moment is zero. In the next step, we terminate O ions close to reduced Ni by H. We find that $Ni^{3+}$ ions go back to $Ni^{2+}$ in AFM order. It should be noted that only for the NiO pure case, the system is an insulator, i.e., the Fermi energy does not penetrate to the CB or VB.

We next passivate the double–Ni-vacancy configuration with two H atoms, each adsorbed on top of an O ligand that is coordinated to one of the $Ni^{3+}$ centers. This operation aims to emulate the hydroxylated NiO (h-NiO) produced experimentally when $H_2O$ was introduced in the plasma discharge during deposition. Moreover, as confirmed later by TEM analysis, the hydroxylated films preserve the underlying rock-salt NiO structure. The resulting magnetic moments show that most Ni ions still carry $\pm 1.61\mu_B$, as in pristine NiO, but the strongly reduced Ni sites associated with the vacancies are no longer present. The moments on these Ni ions increase from ± 0.85$\mu_B$ in the bare double-vacancy case to values close to the bulk (± 1.58$\mu_B$). Instead, a few Ni ions exhibit only moderately reduced moments of about 1.47–1.55 $\mu_B$, indicating that the holes are now more delocalized over the Ni–O–H network rather than being localized on specific $Ni^{3+}$ sites. Some O ions again acquire finite spin polarizations up to 0.06 $\mu_B$.

In contrast, the H atoms carry only very small moments of opposite sign (±0.003 $\mu_B$), so that the total magnetization of the supercell remains essentially zero. The spin-resolved DOS (Figure 1d) shows that the Fermi energy is pinned at the VBM, and a sharp, almost spin-degenerate H-1$s$ peak appears as a deep, localized state well below the valence band, together with a weak H contribution near the top of the VB, reflecting hybridization with O-2$p$ states. Thus, in this local defect configuration, H incorporation passivates the $Ni^{3+}$ centers created by the Ni vacancies, effectively driving them back toward a $Ni^{2+}$-like configuration at the hydroxylated sites.

To rationalize these electronic reconstructions within a chemical framework, the DFT-predicted charge redistributions can be expressed through a corresponding defect-compensation reaction that preserves electroneutrality after plasma growth. These mechanisms determine how Ni vacancies, oxygen acceptor states, and hydroxyl species distribute charge within the lattice under different plasma conditions and are summarized in Equations 1–4.

Under $O_2$-rich sputtering conditions (no $H_2O$), NiO deviates from stoichiometry through the formation of Ni vacancies ($V_{Ni}^{''}$). The DFT calculations show that isolated vacancies are primarily compensated by the formation of oxygen-ligand holes ($O^-$), which appear as shallow unoccupied states just above the valence band maximum. At higher vacancy concentrations, neighboring vacancies interact cooperatively, oxidizing adjacent $Ni^{2+}$ to $Ni^{3+}$ and forming localized $Ni^{3+}$–$O^-$ pairs. At sufficiently high oxygen chemical potential, incorporation of oxygen interstitials is also possible, although not seen in our experiments and calculations. Together, these vacancy-driven mechanisms generate the acceptor states and ligand-hole configurations identified in the DOS analysis, accounting for the p-type character of Ni-deficient NiO[34–37].

When $H_2O$ vapor is introduced during plasma deposition, hydroxyl species bind preferentially to O sites adjacent to Ni vacancies. The DFT results show that $OH^-$ incorporation compensates for the negative charge of $V_{Ni}^{''}$, and restores nearby $Ni^{3+}$ centers to $Ni^{2+}$, removing the di-vacancy in-gap states seen in Figure 1c. The resulting electronic structure contains only shallow OH-derived states (Figure 1d), arising from mixed H 1$s$–O 2$p$–Ni 3$d$ character, and does not stabilize $Ni^{3+}$–$O^-$ ligand-hole configurations. This hydroxylation removes the vacancy-associated deep defect states rather than generating $Ni^{3+}/O^-$ ligand-hole configurations, and thus produces the shallow unoccupied state seen in Figure 1d.

Together, these mechanisms highlight that controlling the $O_2/H_2O$ ratio during plasma growth directly tunes Ni-vacancy formation and Ni–O–H covalency, which in turn influence hole delocalization. To validate these

DFT-based predictions experimentally, we synthesized NiO and hydroxylated NiO (h-NiO) thin films under systematically varied plasma compositions, as described in the following section.

2.2 Synthesis and physical characterization of NiO and h-NiO thin films

NiO and h-NiO thin films were deposited via reactive magnetron sputtering onto polished nickel foils. For the preparation of NiO thin films, oxygen content in the plasma discharge was used at 1.0%, 20%, 66%, and 88%. For the preparation of h-NiO thin films, water vapor was introduced in the $1.0 \times 10^{-6}$ to $1.0 \times 10^{-4}$ mbar partial pressure range. The deposition parameters used for NiO and h-NiO films are summarized in Tables S1, S2, and S3. For h-NiO thin films and high $O_2$ content in the plasma discharge, the maximum thickness was limited to approximately 15 nm due to a significant reduction in deposition rate. This reduction originates from the reaction of water vapor with the target erosion zone, which forms a less conductive nickel-oxide layer on the target surface and substantially reduces the sputtering yield. This layer causes a substantial decrease in target voltage, thereby reducing the deposition rate[38]. However, two reference samples were prepared with 1.0% and 20% $O_2$ content, having a thickness of 100 nm, to obtain an acceptable sample for Grazing Incidence X-Ray Diffraction (GXRD) measurements.

The morphology of the 15 nm and 100 nm films prepared with 1.0%, 20%, 66%, and 88% $O_2$ + $H_2O$ mixtures in the plasma discharge is presented in the 3D AFM micrographs in Figure S1. Two scan sizes ($10 \times 10$ µm$^2$ and $0.5 \times 0.5$ µm$^2$) were used to capture distinct morphological features, with the larger scan highlighting the macroscopically flat regions of the samples, for NiO prepared with 1.0% $O_2$ content (Figure S1a) the surface roughness increases with film thickness, characterized by a surface covered with particles exhibiting a pyramidal-shaped structure and a dendritic microstructure (Figure S1e). In contrast, higher oxygen concentrations progressively reduce surface roughness, as observed in the 15 nm films (Figure S1a-c) and in the 100 nm thickness film (Figure S1f). The variation in surface morphology and roughness can be attributed to increased defect formation during growth. At higher oxygen partial pressure, the amount of metallic $Ni^0$ decreases, likely due to enhanced oxidation of sputtered Ni atoms and changes in grain-growth kinetics.[38,39]

The GXRD diffractogram of the 100 nm thickness NiO sample prepared with 1.0% $O_2$ (Figure S2) displays four NiO Bragg reflections at (111), (200), (220), and (311) in line with reported values[40,41], together with three reflections from the Ni substrate at (111), (200), and (220). Among these, the NiO (200) peak at $2\theta = 43.9°$ is particularly intense, indicating preferential growth along this crystallographic direction. This preferential (200) orientation reflects the low surface energy of the nonpolar (100)/(200) planes in NiO, which are more stable than the polar (111) surface[11]. The additional reflection at $2\theta = 44.6°$ corresponds to the Ni (111) lattice, originating from the underlying substrate.

As the $O_2$ concentration in the plasma process increases, the intensity of the NiO (200) reflection decreases significantly, indicating a shift in the preferential orientation from (200) to (111). This trend is attributed primarily to the reduced adatom mobility under oxygen-rich sputtering conditions[39,42]. Higher $O_2$ flow increases scattering and promotes partial oxidation of sputtered species, both of which reduce surface diffusion and favor growth along the (111) direction[39].

Electrochemical conditioning further modifies the structural orientation. After 200 CV cycles, the 100 nm NiO film deposited at 1.0% $O_2$, transforms at the surface into hydrous $Ni(OH)_2$ (α-$Ni(OH)_2$) consistent with previous reports[10,17], this transformation coincides with the disappearance of the intense (200) reflection, suggesting that electrochemical conditioning leads to partial loss of the (200) texture, consistent with surface reconstruction during NiO → $Ni(OH)_2$ transformation. The absence of distinct $Ni(OH)_2$ reflections in the XRD pattern is likely due to the poor crystallinity or low volume fraction.

The structural response to conditioning depends strongly on the initial defect density and preferred orientation established during deposition. For the 20% $O_2$ film, which initially exhibits a stronger (111) texture and higher Ni deficiency, electrochemical conditioning drives the opposite evolution: the intensity of the (111) reflection decreases significantly. In contrast, a pronounced (200) reflection emerges, suggesting that the initially exposed (111) surface reacts with $OH^-$ ions during conditioning and undergoes structural reconstruction toward a (200)-like texture, consistent with the preference for nonpolar surfaces during hydroxylation. A higher concentration of Ni deficiencies in the film likely facilitates this rearrangemen[43].

Repeated redox cycling during conditioning enhances ionic mobility and promotes structural reorganization within the film, potentially modifying an initially textured structure into a more polycrystalline or less oriented arrangement[44]. As a result, the observed decrease in the (111) reflection and increase in the (200) reflection capture the dynamic structural evolution of NiO under electrochemical stress.

While both Ni and NiO crystallize in a face-centered cubic (fcc) structure, their different lattice parameters result in Bragg reflections of the same family of planes appearing at different 2θ angles. For instance, the (111) reflection occurs at 37.11° for NiO and 44.64° for Ni, reflecting the intrinsic lattice mismatch between NiO and the Ni substrate, which results in residual strain within the film[45]. Increasing the $O_2$ concentration shifts the 2θ position and broadens the full width at half maximum (FWHM), consistent with reduced coherent domain size as inferred from the Scherrer equation[46].

The broadening arises because, when the crystallite size approaches the X-ray wavelength (λ), fewer coherent scattering planes contribute to the diffraction, resulting in weaker and broader peaks[47]. However, given the low signal-to-noise ratio in the NiO diffraction patterns and the high defect density of the films, the observed FWHM broadening likely reflects a combination of structural disorder and limited crystallite size[48]. In such cases, applying the Scherrer equation would yield inaccurate estimates of crystallite size.

In contrast to the clearly separated peaks of Ni and NiO arising from their distinct lattice parameters, although Gong et al.[22] attributed the reflections at 2θ = 43.1°, 62.6°, and 75.2° to the $Ni_2O_3$ phase, such assignments are controversial because these peaks overlap with Ni and NiO reflections. Therefore, we cannot give clear evidence to support the presence of $Ni_2O_3$ in our samples. Consequently, complementary spectroscopic analysis would be required to unambiguously resolve the chemical composition and oxidation states of the films.

The evolution of preferred orientation from (200) to (111) with increasing $O_2$ content, together with peak broadening and subtle 2θ shifts, points to lattice contraction and enhanced disorder typical of Ni-deficient $NiO_x$[38,39]. These structural fingerprints indicate that oxygen-rich plasma conditions promote Ni-vacancy formation and alter the local bonding environment. To examine how plasma composition modifies Ni/O stoichiometry, oxidation-state distribution, and spectroscopic signatures associated with ligand-hole formation, we performed in-vacuum XPS analysis, as detailed in the following section.

## 2.3 Chemical Composition Analysis

In-system XPS measurements were conducted on the as-prepared thin films to determine their elemental composition and nickel oxidation states. The quantitative Ni/O atomic ratios derived from the survey spectra are shown in Figure S3. NiO and h-NiO samples deposited at higher $O_2$% concentrations exhibit a clear Ni deficiency, except for the film grown at 1.0% $O_2$. Although the reduced Ni/O ratio does not imply the presence of well-defined, isolated Ni vacancies, it reflects a net loss of Ni from the lattice, which can be formally represented as cation-site vacancies ($V_{Ni}^{''}$) in Kröger-Vink notation. Interestingly, the overall oxygen content does not scale linearly with the $O_2$ fraction in the plasma, suggesting nonlinear incorporation dynamics during growth.

Under the highly oxidizing plasma conditions used here, the formation of oxygen vacancies is thermodynamically disfavored, and any such sites would rapidly re-oxidize upon exposure to residual oxygen or moisture[49,50], since reactive oxygen species are abundant in both the plasma phase and at the film surface. Additionally, XPS is not capable of directly probing oxygen vacancies because such sites heal spontaneously upon exposure to oxygen or moisture. Therefore, the O 1s component at ~531 eV –frequently incorrectly ascribed to oxygen vacancies in the literature– arises instead from hydroxyl groups, adsorbed oxygen, or other surface species, as emphasized by Idriss[49] and reinforced by Easton and Morgan[50]. Therefore, the observed nonstoichiometry is attributed primarily to Ni-site vacancies rather than oxygen deficiency, consistent with defect-chemistry models of NiO under O-rich conditions[51].

Ni deficiency is not merely a structural feature but directly affects the electronic properties of the films. Vacancy defects are known to introduce electronic states, modify the band structure, and shift the work function, thereby influencing catalytic activity[52,53]. For instance, Mark et al.[54] modified the work function of NiO by introducing oxygen vacancies, which, as a result, moved the Fermi level away from the VBM. Similarly, Egbo et al.[55] demonstrated through spectroscopic ellipsometry and first-principles calculations that Ni vacancies alter both the electronic and optical responses, highlighting their role in adjusting conductivity and reactivity. By analogy, the Ni deficiency observed in our sputtered films is expected to contribute not only to the chemical composition, as evident in the XPS features, but also to the catalytic performance discussed in Section 2.5.

The compositional trend at high $O_2$ partial pressures (Figure S3) can be attributed to "target poisoning", where the formation of NiO on the target surface reduces the sputtering yield of Ni, thereby lowering the Ni flux to the substrate[56,57]. Meanwhile, reactive oxygen species remain abundant, yielding a Ni-deficient film composition. Such changes in the $O_2$ partial pressures in the plasma discharge not only impact elemental fluxes but also reflect broader shifts in growth dynamics under increasingly oxidizing conditions, as already evidenced in Section 2.2.

XPS analysis of the Ni $2p_{3/2}$ and O $1s$ core levels for films prepared at different oxygen partial pressures is presented in Figure 2a-c, together with the spectrum of a film deposited in the presence of 0.3% $H_2O$ for comparison. The O $1s$ spectra (Figure 2a and c) consistently show a main peak at ~529.0 eV, corresponding to lattice oxygen ($O^{2-}$), and a higher binding energy shoulder at ~530.6-531.0 eV, which becomes more intense with increasing $O_2$ concentration. This feature is often attributed to non-lattice oxygen species such as adsorbed oxygen or oxidized lattice oxygen ($O^-$). It may also reflect the presence of defect-related species as Ni vacancies or oxygen bonded to $Ni^{3+}$ centers, which contribute to defect states in the oxide network[34,58–60], and hydroxyl groups for films prepared with $H_2O$.

Interestingly, as the oxygen concentration in the plasma increases beyond 20%, the entire O $1s$ envelope shifts toward higher binding energies. In many oxide systems, such as $TiO_2$ or $CeO_2$, increasing $O_2$ during deposition typically shifts the O $1s$ peak to lower binding energies due to vacancy passivation[61,62]. In NiO, however, the opposite trend occurs; the O $1s$ peak shifts to higher binding energy. This shift has been attributed to enhanced Ni-vacancy formation and reduced final-state screening under highly oxidizing conditions[63]. When $H_2O$ is added to the plasma deposition environment, the peak at ~530.6 eV broadens and increases, consistent with the formation of surface hydroxyls[49], indicating that reactive sputtering in the presence of water vapor promotes hydroxylation rather than further oxidation towards higher nickel valence states. The feature at ~532.1–532.4 eV corresponds to molecularly adsorbed $H_2O$, whereas the peak at ~535.5 eV originates from gas-phase $H_2O$[64,65].

Figure 2b shows the Ni $2p_{3/2}$ core level of the films. The dominant peak at ~854 eV is assigned to the well-screened Ni $2p^53d^9Z$ state, while the shoulder at ~855 eV corresponds to the Ni $2p^53d^9L$ state associated with O $2p$ ligand holes[66]. The distinction between these two features has been a subject of debate. Stevie et al.,[67] attributed the doublet at 853–856 eV to the multiplet splitting of the Ni $2p_{3/2}$ core level. Regardless of the interpretation, the increase in $O_2$ concentration enhances the Ni $2p^53d^9L$ state (~855 eV), indicating a higher density of oxygen ligand-holes ($O^-$) states, in agreement with the O $1s$ analysis. The shake-up line at ~861 eV arises from de-excitation processes, in which the outgoing photoelectrons excite Ni $3d$ electrons, thereby reducing their kinetic energy and shifting the peak to higher binding energy[68].

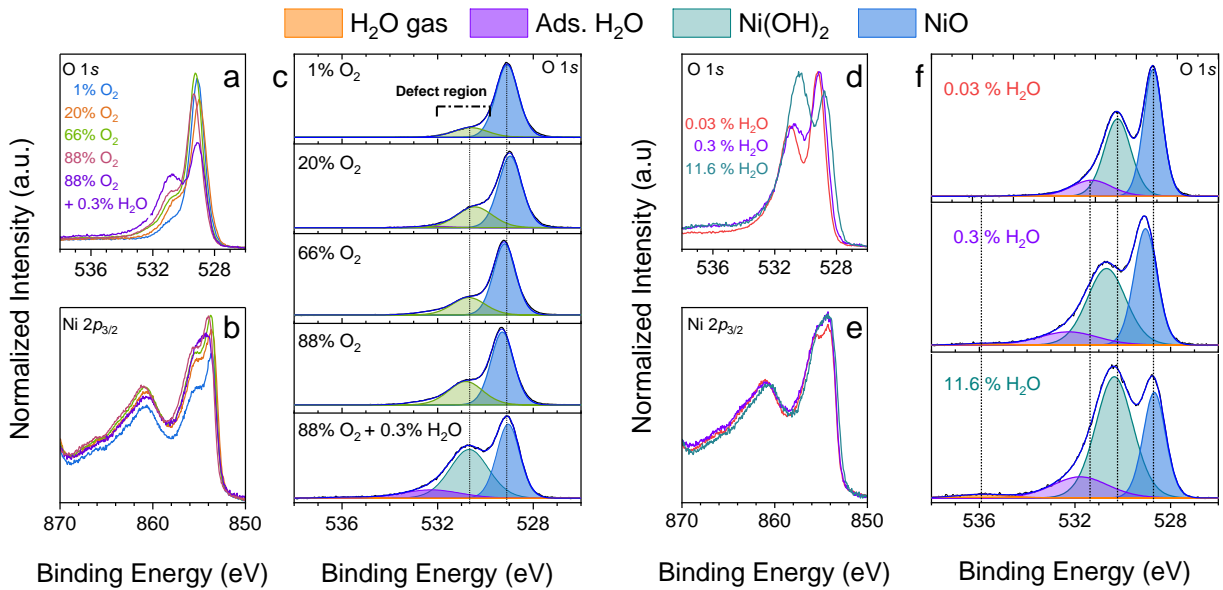

Figure 2. XPS analysis of NiO thin films prepared with different oxygen concentrations and with varying $H_2O$ concentrations in the plasma discharge. (a, b) O $1s$ and Ni $2p_{3/2}$ spectra of thin films of 15 nm deposited with 1.0, 20, 66, and 88% $O_2$, and with 88% $O_2$ + 0.3% $H_2O$ in the plasma discharge. (c) Deconvolution of the O $1s$ spectra from (a), showing contributions from lattice oxygen ($O^{2-}$), hydroxyl groups, adsorbed water, and defect-related species. (d, e) O $1s$ and Ni $2p_{3/2}$ spectra for films deposited with 0.03, 0.3, and 11.6% $H_2O$ in the plasma discharge. (f) Deconvolution of the O $1s$ spectra from (d), showing contributions from lattice $O^2$, hydroxyl species, adsorbed water, and gas-phase $H_2O$. All spectra were collected before air exposure and prior to electrochemical treatment.

When $H_2O$ is introduced during deposition, the evolution of the Ni $2p$ line shape must be interpreted differently. In hydroxylated NiO, changes in the $3d^9L/3d^9Z$ intensity ratio cannot be directly or solely attributed to the formation of additional $Ni^{3+}$–$O^-$ ligand-hole pairs. Instead, they arise from modified final-state screening and stronger Ni–O–H hybridization, which alters the multiplet balance without generating the deep divacancy-derived ligand-hole states characteristic of vacancy-rich NiO. This behavior is fully consistent with DFT predictions that hydroxyl groups compensate vacancy-induced $Ni^{3+}$ centers and suppress deep in-gap states.

Figure 2d and e compare the O $1s$ and Ni $2p_{3/2}$ spectra of films grown at 88% $O_2$ with increasing $H_2O$ partial pressure. The main O $1s$ peak at ~529.1 eV corresponds to lattice $O^{2-}$, while the shoulder near ~530.9 eV is dominated by hydroxyl (OH⁻) species introduced during H$_2$O-assisted sputtering. This region may still contain contributions from oxidized oxygen and other defect-related environments; the higher-energy feature at ~532.1 eV comes from adsorbed water, as we already mentioned. With increasing H$_2$O content, the hydroxyl component grows at the expense of the NiO lattice contribution, indicating progressive hydroxylation of surface and near-surface regions. In the corresponding Ni 2$p$ spectra, the increase of the $3d^9$L relative to the $3d^9$Z final state indicates globally altered screening and changes in the local Ni–O–H coordination environment. Such an increase is also consistent with a higher degree of ligand-hole character, in line with the Ni L-edge results presented later, which show persistent ligand-hole signatures in hydroxylated films. Thus, the evolution of the Ni 2$p$ multiplet structure reflects a combination of modified screening, increased Ni–O–H coordination, and a redistribution of unoccupied 3$d$–O 2$p$ hybrid states that accompanies hydroxyl incorporation.

This interpretation is further supported by Ni K-edge XANES (Figure 3l). Both NiO and h-NiO display an absorption edge ($E_0$=8346 eV), indicating that the average Ni oxidation state remains unchanged. The pre-edge feature at ~8333.5 eV, arising from 1$s \rightarrow$ 3$d$ transitions, that gains intensity through Ni 3$d$–O 2$p$ hybridization[69–71], is shown for both samples. Such pre-edge is well-established as a signature of increased metal-ligand covalency and oxygen-hole character in transition-metal oxides[72]. The pre-edge intensity observed in both NiO and h-NiO films, therefore, provides complementary evidence of ligand-hole formation (Ni$^{3+}$–O⁻ or O-centered) and covalency in these materials.

To benchmark the chemical and electronic changes induced by H$_2$O incorporation, we compared the O 1$s$, Ni 2$p_{3/2}$, and valence band XPS spectra of three samples (Figure 3a-c): NiO prepared with 1.0% O$_2$, Ni(OH)$_2$, and h-NiO (with 11.6% H$_2$O). The O 1$s$ peak of the reference samples, NiO and Ni(OH)$_2$, appear at ~529.1 eV and ~531.6 eV, respectively, consistent with previously reported values for these Ni oxide phases[16,73]. The corresponding Ni 2$p_{3/2}$ core level for these two samples is located at 853.6 eV and 856.1 eV, respectively, with their VBM located 0.3 eV below the Fermi level, for NiO–reflecting its p-type semiconducting character[74]–and 1.2 eV for Ni(OH)$_2$, consistent with its n-type behavior[6,10].

In contrast, the h-NiO sample shows clear signatures of partial hydroxylation. The O 1$s$ spectrum exhibits contributions from both NiO and Ni(OH)$_2$, with a distinct component at ~530.5 eV assigned to OH⁻ species, and a reduced oxygen lattice signal near 528.8 eV, consistent with partial hydroxylation (Figure 3a, red). Similarly, the Ni 2$p_{3/2}$ core level appears between the positions of the two reference phases, consistent with mixed O$^{2-}$/OH⁻ coordination (Figure 3b, red). The valence band spectrum of h-NiO (Figure 3c red) shows a slightly deeper VBM than pure NiO, with the VBM located ~0.4 eV below the Fermi level. Additionally, the linear extrapolation of the spectral tail approaches the Fermi level (within ~0.07 eV), consistent with defect-related states arising from Ni vacancies and impurity bands[5,55,75].

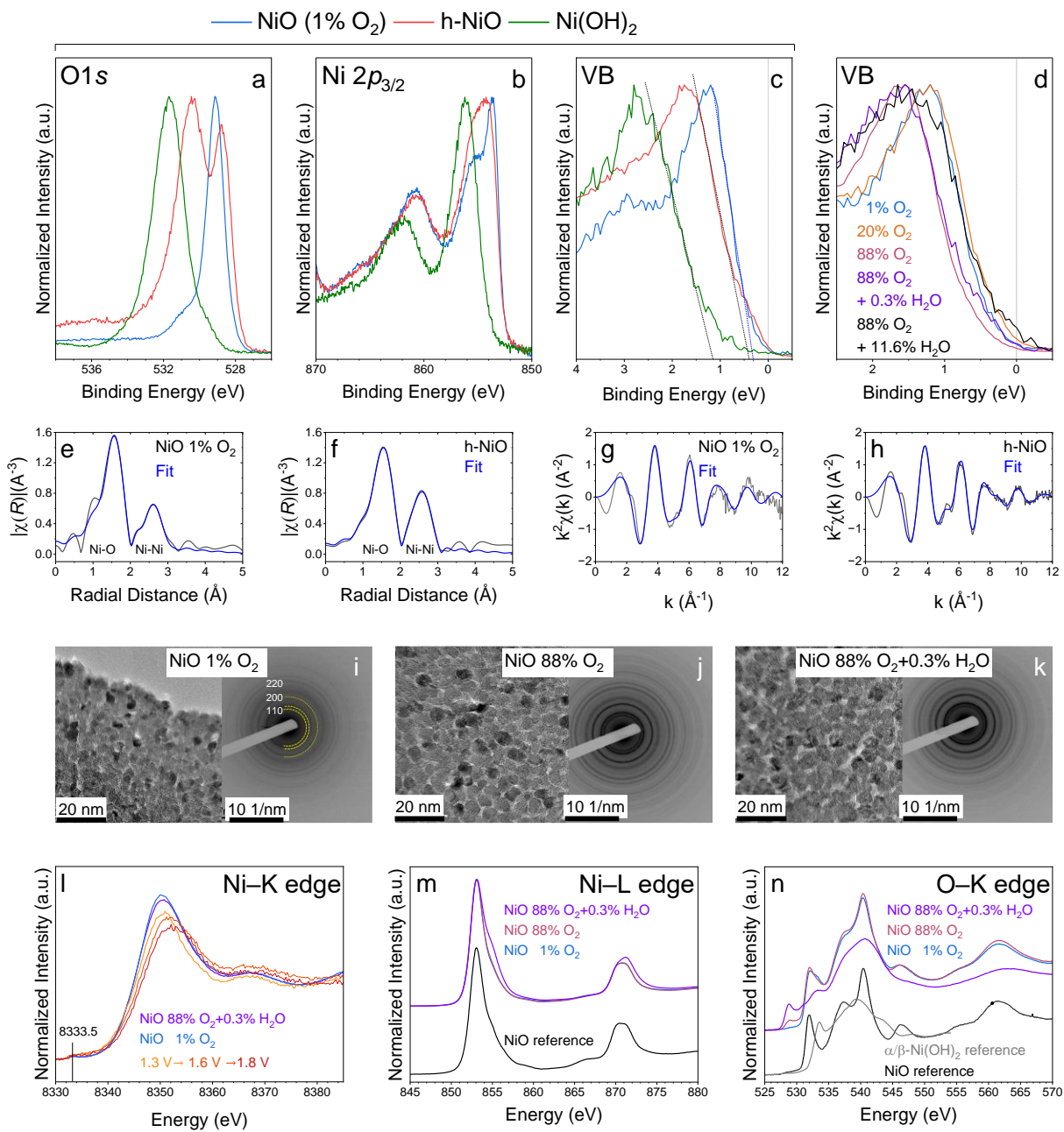

Figure 3. XPS comparison of NiO, Ni(OH)$_2$, and h-NiO thin films. (a) O 1$s$ core-level spectra showing lattice oxygen in NiO, hydroxyl-related species in Ni(OH)$_2$, and mixed contributions in h-NiO. (b) Ni 2$p_{3/2}$ spectra highlighting the intermediate binding energy of h-NiO, consistent with mixed O$^{2-}$/OH$^-$ coordination. (c) Valence band spectra illustrating the intermediate electronic character of h-NiO between NiO and Ni(OH)$_2$. (d) Comparison of VBM for NiO and h-NiO, showing that h-NiO exhibits a reduced VBM and defect-related tail states extending above the Fermi level. (e-h) Magnitude of the Fourier transform of the k$^2$-weighted χ(k) spectra in R-space for NiO (e) and h-NiO (f), showing the first peak from the Ni-O scattering path (~1-2 Å, phase-uncorrected) and the second peak from the Ni-Ni scattering path (~2–3 Å). (g, h) Corresponding k$^2$-weighted χ(k) spectra in k-space for NiO (g) and h-NiO (h), with experimental data and best-fit curves. (i–k) High-resolution TEM and Selected-area electron diffraction (SAED) images of NiO prepared with 1.0% O$_2$ (i), 88% O$_2$ (j), and h-NiO (k). The SAED patterns show cubic NiO diffraction rings for both NiO and h-NiO, confirming preservation of the rock-salt structure. (l) Ni-K edge of ex-situ spectra of NiO and h-NiO, highlighting the pre-edge intensity at ~8333.5 eV, which reflects enhanced Ni 3$d$–O 2$p$ hybridization, and in operando conditions in 1M KOH at three different potentials of 1.3 V, 1.6V, and 1.8V vs RHE, indicating the formation of NiOOH OER active phase. (m,n) Ni-L, and O-K edges for the samples with 1.0 and 88% O$_2$, 88% O$_2$ + 0.3 % H$_2$O. The samples are compared to a NiO reference from Sigma Aldrich and a σ/β-Ni(OH)$_2$ sample.

Although the h-NiO VBM lies slightly deeper than in NiO, the near-edge spectral tail does not necessarily indicate the formation of additional ligand-hole states. Instead, these shallow states could also originate from Ni–O–H hybridization and from the increased local electronic and symmetry disorder introduced by hydroxylation, which broadens the upper valence band. In our DFT model, hydroxylation compensates the Ni-vacancy–induced $Ni^{3+}$ centers and suppresses the deep in-gap states associated with clustered vacancies, restoring a predominantly $Ni^{2+}$ like configuration around the hydroxylated sites. This DFT description reflects the local electronic reconstruction that occurs specifically at hydroxylated vacancies and should not be interpreted as a change in the global Ni oxidation-state distribution of the film. The $Ni^{3+}$ species discussed here correspond to vacancy-induced ligand-hole states in the as-prepared films and are distinct from the NiOOH-type $Ni^{3+}$ generated electrochemically during OER.

The experimental evolution of the VBM is consistent with the sequence of defect configurations predicted by DFT (Section 2.1). At low oxygen concentration (1.0 % $O_2$), the XPS VBM lies further below the Fermi level, corresponding to the stoichiometric or single-vacancy regime– as predicted by DFT– in which only isolated acceptor-like states appear slightly above the VBM. Increasing the $O_2$ fraction (20-66 %) drives the VBM progressively closer to the Fermi level and introduces discernible valence band tailing, in agreement with the DFT description of two-vacancy configurations that generate partially filled O 2$p$–Ni 3$d$ hybrid states near the Fermi level. At the highest $O_2$ content (88 %) and under $H_2O$ conditions, the VBM slightly shifts away from the Fermi level. This apparent VBM upshift reflects defect-induced broadening rather than a rigid movement of a true band edge. In the hydroxylated film, the valence-band onset becomes even less well defined. Enhanced Ni–O–H hybridization broadens the upper valence band through enhanced Ni–O–H hybridization and strengthens band tailing, causing the extrapolated VBM to appear deeper.

## 2.4 Local Structural Environment from Ni K-edge EXAFS

To further probe the short-range structure and defect chemistry of NiO and h-NiO, Ni K-edge (~8333.5 eV), an EXAFS analysis was performed on the as-deposited films before electrochemical conditioning (Figure 3e-h). The Fourier-transformed spectra and corresponding fits provide direct insight into local coordination, bond distances, and structural disorder.

In both cases, the structural model that best describes the local environment is the cubic rock-salt structure. This structural model is well established for stoichiometric NiO, which exhibits a well-defined $NiO_6$ octahedral environment with highly reproducible Ni–O (~2.08 Å) and Ni–Ni (~2.95–3.00 Å) distances reported in neutron diffraction and X-ray absorption studies[71,76,77], providing a physically justified reference for EXAFS modeling. To better track the evolution of the system following the treatment, the same parameters were refined for both samples: the edge energy shift, the scattering amplitude, the interatomic distances, and the Debye–Waller factors for the Ni–O and Ni–Ni shells

For NiO thin film (Figure 3e and g), the fitted Ni–O bond length of 2.056 ± 0.006 Å agrees well with these literature values, while the Debye–Waller factor ($\sigma^2$ = 0.0060 ± 0.0009 Å$^2$)[71] lies near the upper end of reported bulk values (~0.004 Å$^2$). This increased disorder may arise from the higher surface-induced strain in thin films compared to the bulk material, broadening the first-shell distance distribution.

The Ni–Ni shell appears at 3.05 ± 0.01 Å, a value slightly larger than reported. Moreover, the pronounced local disorder ($\sigma^2$ = 0.024 ± 0.002 Å$^2$) is substantially higher than bulk references (~0.010–0.015 Å$^2$)[71]; this behavior mirrors the trend observed in the first shell, supporting the idea of structural distortion driven by significant surface strain contributions[76–78]. These parameters reproduce both the amplitude and phase of the experimental first two shells, indicating that while the local NiO$_6$ octahedron is maintained, the cation-cation environment is more significantly distorted.

In addition, the weak but discernible pre-edge feature in the XANES region at ~8333.5 eV (Figure 3l) reflects enhanced Ni 3$d$–O 2$p$ hybridization and deviations from centrosymmetric octahedral coordination. Increased pre-edge intensity can arise from several mechanisms–including vacancy-induced distortions, local symmetry breaking, and ligand-hole character–and thus cannot be attributed exclusively to Ni vacancies. Although our EXAFS cannot quantify vacancy concentration directly, the combination of higher $\sigma^2$ values and the enhanced pre-edge feature in NiO points to increased local disorder and stronger Ni 3$d$–O 2$p$ hybridization relative to bulk NiO.

Regarding the h-NiO sample (Figure 3f and h), the first-shell Ni–O distance is 2.056 ± 0.007 Å with $\sigma^2$ = 0.007 ± 0.001 Å$^2$. This Ni–O distance is essentially identical to that of NiO within the error bars, indicating that the H$_2$O treatment does not induce any measurable structural modification in the first coordination shell. The Ni–Ni shell appears at 3.02 ± 0.01 Å, with $\sigma^2$ = 0.020 ± 0.001 Å$^2$: after the treatment, this second shell becomes slightly shorter and shows reduced second-shell structural disorder (i.e., it shows smaller interatomic distances and lower $\sigma^2$). Although the Ni–Ni distance and $\sigma^2$ values remain higher than bulk NiO, the reduction in both parameters compared to the oxygen-rich NiO film indicates a partial relaxation of the medium-range structure upon hydroxylation, suggesting that h-NiO retains thin-film–induced strain but exhibits reduced vacancy-induced distortion relative to the 88% O$_2$ sample.

A comparison of the edge energy position and scattering amplitude between the two samples provides insight into the evolution of the absorber oxidation state and the possible formation of defects. After refinement, the edge energy is 8344.4 ± 0.6 eV (8344.38034 ± 0.55777) for both samples (8344.38034 ± 0.55777 in NiO and 8344.43815 ± 0.61934 in h-NiO), clearly indicating that the bulk average oxidation state remains unchanged. The $S_0^2$ value also remains constant: it refines to 0.86 ± 0.04 (0.8633 ± 0.0447) for NiO and 0.86 ± 0.05 (0.86228 ± 0.05034) for h-NiO, indicating that the overall amplitude reduction and the fraction of coherently scattering atoms remain comparable between the two samples. Because $S_0^2$ is primarily a global scaling factor[79], its similarity does not rule out local structural modifications. When combined with the changes in Ni–Ni distance and $\sigma^2$, the EXAFS results instead indicate that hydroxylation induces a partial relaxation of vacancy-induced distortions without generating a fundamentally different short-range coordination environment.

Complementary TEM analysis (Figure 3i-k) supports these structural trends. Selected-area electron diffraction (SAED) patterns show that films grown at low and high $O_2$ content, as well as with $H_2O$ content, retain the cubic rock-salt NiO phase without additional reflections.

The Ni–Ni distances and coordination numbers differ from those of layered $Ni(OH)_2$, confirming that hydroxylation modifies only the first coordination shell while leaving the medium-range NiO lattice intact. This interpretation is consistent with the DFT results, which show that H incorporation locally perturbs the vacancy environment, reduces the $Ni^{3+}$ character around hydroxylated vacancies, and removes the deep divacancy-related in-gap states without reconstructing the octahedral framework. Importantly, this local passivation mechanism does not imply the full elimination of $Ni^{3+}$ species in the real film, as experimentally observed by Ni L-edge(Figure 3m); instead, it reflects the electronic restructuring that occurs specifically at hydroxylated sites. Thus, the structural response to hydroxylation is best described as localized Ni–O–H disorder embedded within a largely preserved NiO lattice, rather than a phase transition toward bulk $Ni(OH)_2$.

Operando Ni K-edge XANES of the h-NiO film (Fig. 3l) demonstrates that the material undergoes a clear, potential-driven oxidation even without prior conditioning. Holding the electrode for 30 min at 1.3 V produces a spectrum characteristic of predominantly $Ni^{2+}$ in a hydroxide-like environment. Increasing the potential to 1.6 V for another 30 min produces a distinct positive edge shift, marking the onset of $Ni^{3+}$ formation. At 1.8 V for 30 min, the edge shift grows to ~1.9 eV relative to the as-prepared h-NiO sample, consistent with the establishment of the γ-NiOOH electronic structure. Notably, this potential-dependent evolution is significantly sharper than in our previous operando study[10], indicating that predefined Ni–O–H environments facilitate a more efficient structural and electronic reorganization under anodic polarization.

Beyond the Ni K-edge analysis, complementary O K-edge and Ni L-edge X-ray absorption measurements provide further insight into how oxygen-rich and hydroxylation modulate the electronic structure of NiO and h-NiO. Together, these absorption edges govern transitions into the unoccupied 3$d$ electronic states that are essential for understanding OER catalysis.

The O K-edge spectra of the three representative samples (Figure 3n) show how vacancies and hydroxylation modify the unoccupied O 2$p$–Ni 3$d$ manifold. The stoichiometric-like 1% $O_2$ film closely resembles the NiO reference, with a weak pre-edge at ~529 eV, consistent with a predominantly $Ni^{2+}$–$O^{2-}$ octahedral environment and limited hole density. This NiO-like configuration is further supported by its Ni L-edge spectrum, which also matches the reference.

In the 88% $O_2$ film, the pre-edge intensity increases and broadens toward ~529–530 eV. Similar enhancements in Ni oxides have been associated with ligand-hole formation driven by Ni vacancies ($Ni^{3+}$–$O^-$ configurations) and enhanced Ni $3d$–O $2p$ covalency[80,81]. However, in our case, the Ni L-edge does not show sharply resolved $Ni^{3+}$ multiplet features, indicating that any $Ni^{3+}$ population remains below the sensitivity of this measurement or is partially delocalized, consistent with our DFT results, which show that isolated vacancies introduce delocalized holes rather than fully localized $Ni^{3+}$ centers. In contrast, higher vacancy densities enhance hybridization and redistribute $3d/2p$ spectral weight near the valence band edge. Thus, the increased pre-edge intensity in the 88% $O_2$ film reflects a combination of vacancy-induced Ni $3d$–O $2p$ hybridization, partially delocalized ligand-hole character, and local symmetry distortions around vacancies, with none of these contributions dominating exclusively.

The hydroxylated h-NiO film (h-NiO, 88% $O_2$ + 0.3% $H_2O$) also displays an enhanced pre-edge, but with subtle differences in line shape. DFT predicts that OH groups bind preferentially at vacancy-adjacent O sites, partially compensating $Ni^{3+}$-like environments and suppressing the deep in-gap states present in the divacancy configuration, restoring more $Ni^{2+}$-like character locally while introducing additional Ni–O–H hybridization in the unoccupied states. Experimentally, this is reflected in the O K-edge by a preserved pre-edge intensity (indicating that vacancy-induced hybridization is not eliminated) and by the Ni L-edge, which shows an increased shoulder near ~855.5 eV and increased $L_2$-edge intensity. These features indicate that ligand-hole character persists at non-hydroxylated regions even as hydroxylation locally alters the vacancy environment.

Together, the O K-edge, Ni L-edge, and DFT results indicate a heterogeneous defect landscape: oxygen-rich growth enhances Ni $3d$–O $2p$ hybridization and generates partially delocalized ligand-hole states; hydroxylation locally compensates some vacancy-induced $Ni^{3+}$-like configurations while preserving significant hybridization in the unoccupied manifold. The observed pre-edge evolution across the series, therefore, reflects variations in covalency, symmetry distortion, and vacancy compensation mechanisms, rather than mutually exclusive spectral origins.

Plasma diagnostics (Figure S4) by optical emission spectroscopy (OES) provide context for these structural trends. The Ar/$O_2$ discharge used for NiO deposition exhibits $Ar^+$ and oxygen-ion emission features[82,83], consistent with an oxidizing environment that favors Ni-vacancy formation. In contrast, the Ar/$O_2$/$H_2O$ plasma contains additional H emission signatures associated with partial water dissociation[84], aligning with the incorporation of Ni–O–H coordination environments detected by XPS. Although OES does not resolve all reactive fragments formed in the discharge, the distinction between oxygen-rich and H-containing plasmas corresponds directly to the defect chemistries extracted from DFT and spectroscopy–vacancy-rich growth under Ar/O$_2$ and hydroxylated surfaces under Ar/O$_2$/H$_2$O. Full plasma diagnostic analysis is provided in the Supplementary Information.

It is important to note that the apparent differences between XRD, EXAFS, and TEM reflect the distinct length scales probed by each technique. XRD reveals substantial modifications in crystallite size, texture, and long-range ordering as a function of plasma composition, whereas EXAFS shows that the local NiO$^6$ octahedral environment remains largely preserved. TEM/SAED confirms this picture: the rock-salt phase persists, although medium-range disorder and strain–clearly visible in XRD peak broadening–are not resolved in SAED. Thus, the films undergo significant long-range structural evolution without altering their fundamental local NiO coordination.

## 2.5 Electrochemical Analysis

The oxygen evolution activity of NiO and hydroxylated NiO (h-NiO) thin films was systematically compared as a function of O$_2$ and H$_2$O content during deposition (Figure 4). In previous studies[5,17,10,16] we have used extensive electrochemical cycling (>1000 CV cycles) to convert the surface into Ni(OH)$_2$, facilitating the formation of the subsequently NiOOH. In contrast, the intentional incorporation of H$_2$O during plasma deposition introduces hydroxyl groups into the films, reducing the need for such long conditioning. Here, only 200 CV cycles between 1.0-1.5 V (vs RHE) were sufficient to precondition the surface for the samples studied, followed by 60 min chronoamperometry (CA) at 1.5 V (vs RHE) to stabilize NiOOH as the active phase in OER. The overpotential at 10 mA cm$^{-2}$, Tafel slopes, and turnover frequency (TOF) were used as descriptors of intrinsic activity and correlated with the defect chemistry and electronic structure modifications described in Sections 2.2–2.4.

### Effect of the Oxygen and H$_2$O Concentration in the plasma discharge on the OER.

NiO films grown under low O$_2$ content (1.0%) exhibit higher overpotentials and a larger Tafel slope (34.5 mV dec$^{-1}$) (Figure 4a-c), which we attribute to the absence of pre-stabilized Ni$^{3+}$/O$^-$ ligand-hole states, following our rationale established by our previous observations and calculations. Increasing the O$_2$ fraction to 20% significantly reduces the slope to 31.4 mV dec$^{-1}$, reflecting slightly accelerated kinetics, which we correlate to higher Ni vacancy densities and shallow acceptor states that enhance Ni–O covalency and shift the valence band closer to the Fermi level (Sections 2.3-2.4). At 66% O$_2$, the slope is further decreased (16.8 mV dec$^{-1}$), indicating faster kinetics, whereas further increasing the O$_2$ content to 88% reverses the performance to 19.8 mV dec$^{-1}$. Since lower Tafel slopes correspond to faster intrinsic reaction kinetics[85–87], these results highlight that an intermediate O$_2$ regime (20-66%) was optimal for stabilizing vacancy-derived ligand-hole character and enhanced Ni–O covalency, which facilitates improved OER kinetics. Similar behavior has been reported in NiO thin films, where excessive oxygen incorporation introduces a high defect density, enhancing carrier scattering and reducing hole mobility, ultimately increasing resistivity and degrading the performance of NiO films[88].

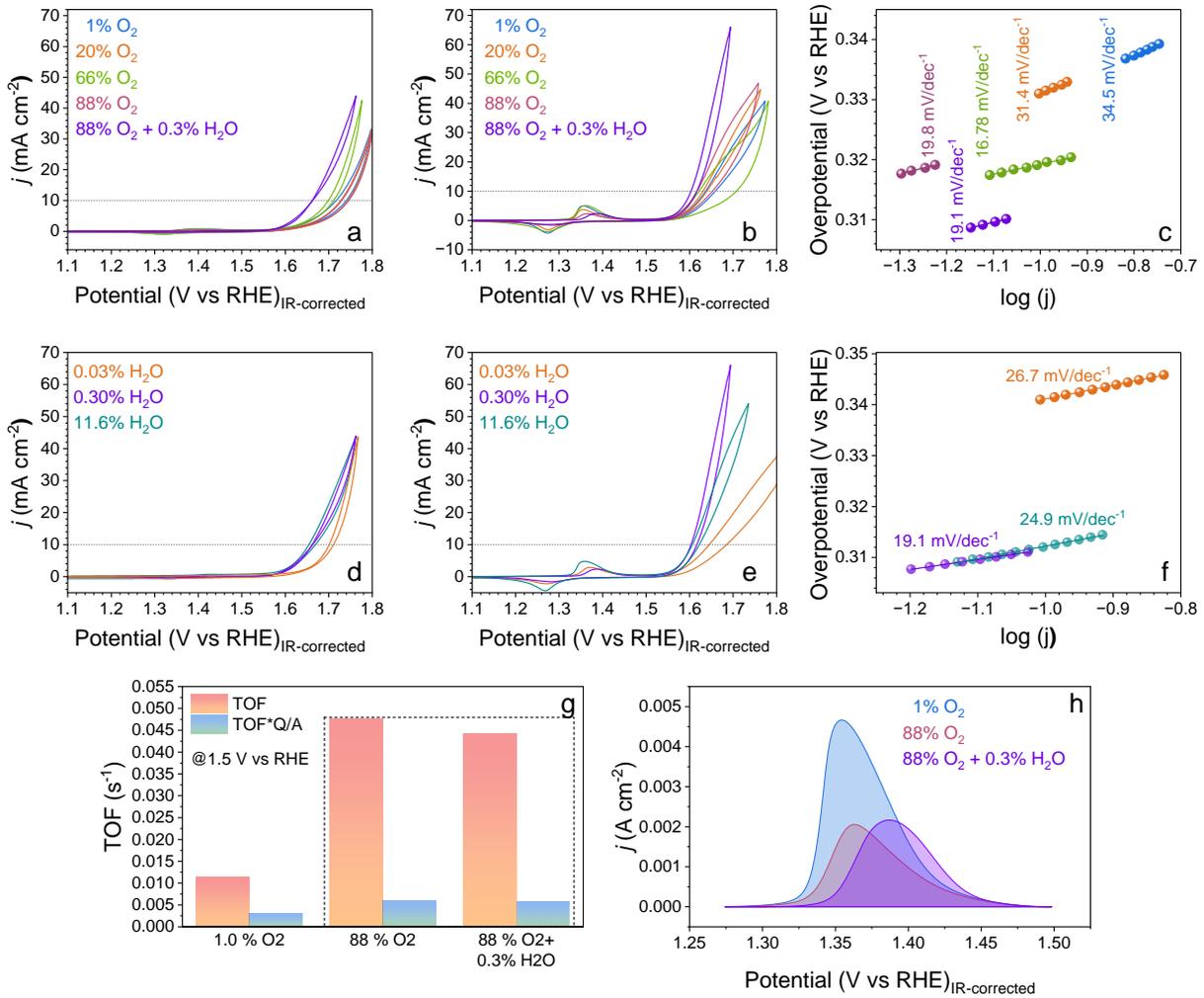

Figure 4 OER activity of NiO thin films as a function of $O_2$ and $H_2O$ content during deposition, highlighting that moderate hydroxylation facilitates the NiOOH formation and optimizes reaction kinetics, while excessive hydroxylation reverses performance. (a) Polarization curves for films deposited under different $O_2$ concentrations (1.0%, 20%, 66%, 88%) and with 0.3% $H_2O$. (b) Polarization curves of the same films after extended conditioning (200 CV cycles between 1.0-1.5 V vs RHE + 60 min chronoamperometry at 1.5 V), where NiOOH is stabilized on the surface. (c) Tafel plots derived from (b), showing optimal activity at intermediate $O_2$ levels and further enhancement by hydroxylation (slope ~19.1 mV dec$^{-1}$). (d–f) Effect of $H_2O$ incorporation (0.03%, 0.3%, 11.6%), where moderate hydroxylation improves OER activity, while excessive hydroxylation decreases performance. (g) TOF and TOF × (Q/A) values extracted at 1.5 V vs RHE for representative samples (1.0 % $O_2$, 88 % $O_2$, 88 % $O_2$ +0.3% $H_2O$). The 88% $O_2$ sample exhibits the highest nominal TOF. The hydroxylated film shows a slightly higher redox charge (Q/A) but a marginally lower TOF×(Q/A), indicating that hydroxylation increases the number of electrochemically addressable Ni sites without improving their intrinsic turnover. Nevertheless, the hydroxylated sample achieves the fastest per-site kinetics, reflected by its lowest Tafel slope. (h) Isolated $Ni^{2+}/Ni^{3+}$ oxidation wave used for charge integration (Q) to determine the electrochemically active Ni population.

The introduction of H₂O vapor during deposition provided an additional pathway to tailor the OER activity. For the samples studied, films grown with 0.3% H₂O exhibited the highest OER activity, with a Tafel slope of 19.1 mV dec$^{-1}$ and reduced overpotentials compared to oxygen-only samples (Figure 4c). Since the electrochemical surface area decreases with H₂O addition (Figure S5)– compared to its counterpart sample prepared only with 88% O₂– this improvement cannot be ascribed to geometric effects but rather to changes in the electronic structure. Moderate hydroxylation introduces mixed Ni–O–H coordination environments and increases first-shell disorder without significantly altering the average Ni–O bond distance (Section 2.4). These local structural perturbations generate shallow Ni–O–H–derived tail states near the valence band edge (Section 2.3), which facilitate charge transfer and accelerate the NiO → Ni(OH)₂ → NiOOH transition during electrochemical conditioning. At higher H₂O content (11.6%), the Tafel slope increased again to 24.9 mV dec$^{-1}$, consistent with excessive hydroxylation increasing disorder and reducing hole mobility.

Overall, the Tafel slopes of NiO and h-NiO (16-35 mV dec$^{-1}$) fall within the low-slope regime predicted by microkinetic analyses[86] and frequently reported for fast alkaline OER kinetics[87], corresponding to ~RT/3F-~RT/2F. Such low slopes have been associated with lattice-oxygen participation in Ni-based oxides[89], although they are not, by themselves, definitive mechanistic proof. In oxygen-rich films, increased Ni-vacancy concentration promotes stronger Ni–O covalency and O-centered ligand-hole character, consistent with DFT-predicted Ni$^{3+}$–O$^-$ configurations. In contrast, hydroxylated films achieve rapid kinetics not through an increase in the vacancy-induced Ni$^{3+}$–O$^-$ ligand-hole density, which is suppressed by hydroxyl incorporation, but through accelerated NiO → Ni(OH)₂ → NiOOH transformation during conditioning. Thus, nickel vacancies modulate the intrinsic electronic structure, whereas hydroxylation modulates the chemical activation pathway, yielding surfaces that behave as pre-activated catalysts with enhanced OER activity[10,17]. The improved OER performance reflects the interplay between vacancy-induced covalency and hydroxylation-mediated NiOOH formation.

To quantitatively evaluate the intrinsic catalytic activity of NiO and h-NiO, the TOF was calculated following the convention established for Ni-based OER catalysts in alkaline media– in which the number of active sites is derived from the Ni$^{2+}$/Ni$^{3+}$ oxidation charge, obtained by integrating the anodic Ni$^{2+}$ → Ni$^{3+}$ peak, and converting this charge (Q) into moles of electrochemically accessible Ni sites via N$_{site}$ = Q/F, assuming a one-electron Ni$^{2+}$ → Ni$^{3+}$ transition– and each evolved O₂ molecule corresponds to a four-electron process[90,91]. The relationship between the measured current density and the number of electrochemically active Ni sites involved in the OER is expressed by equation 1.

$$TOF = \frac{jA}{4FNi_{site}} \qquad (1)$$

where $j$ is the steady-state current density (A cm$^{-2}$) at 1.5 V vs RHE, A is the geometric area (cm²), F is Faraday's constant, and Ni$_{site}$ is the number of electrochemically active Ni sites estimated from the anodic charge associated with the $Ni^{2+}/Ni^{3+}$ redox transition in the cyclic voltammogram. Because $O_2$ evolution involves the transfer of four electrons, the denominator (4F) converts total charge into mol $O_2$ $s^{-1}$ $cm^{-2}$. To facilitate comparison between samples and to account for variations in the electrochemically accessible Ni population, the TOF was also scaled by the redox charge per geometric area (Q/A). This charge-normalized quantity (TOF × Q/A) reflects the effective turnover rate per unit area by combining the intrinsic per-site activity with the experimentally accessible Ni-site population extracted from the $Ni^{2+}/Ni^{3+}$ redox transition.

The resulting TOF values of 0.011 $s^{-1}$ (1.0 %$O_2$), 0.032 $s^{-1}$ (88 %$O_2$), and 0.012 $s^{-1}$ (88 %$O_2$ + 0.3% $H_2O$) are presented in Figure 4g. As the oxygen concentration increased from 1.0% to 88%, the TOF increased by nearly a factor of four, consistent with the formation of vacancy-derived $Ni^{+3}/O^-$ ligand-hole states under oxygen-rich deposition. As expected, the 1.0% $O_2$ sample shows the lowest intrinsic turnover, whereas the 88 %$O_2$ film exhibits the highest TOF due to its stronger ligand-hole character.

Upon controlled hydroxylation (88% $O_2$ + 0.3% $H_2O$), the nominal TOF decreases slightly relative to the oxygen-rich film; even when the redox charge $Ni^{2+}/Ni^{3+}$ redox charge (Q/A = 1.33×10$^{-4}$ C cm$^{-2}$) is somewhat larger compared to the 88 %$O_2$ sample (1.26×10$^{-4}$ C cm$^{-2}$). Despite this, its charge-normalized activity (TOF×Q/A = 0.00588) remains essentially the same as–and slightly lower than–that of the 88 % $O_2$ film (0.00598), indicating that hydroxylation increases the number of electrochemically addressable Ni sites but does not enhance their intrinsic catalytic utilization. Because both samples display comparable Ni/O ratios (Figure S3), the differences in catalytic behavior do not reflect changes in the overall degree of Ni deficiency, but rather differences in how Ni vacancies are charge-compensated and coordinated locally. In the 88 %$O_2$ film, a larger fraction of vacancies is compensated electronically via $Ni^{3+}$ formation and O-centered ligand holes, stabilizing highly active $Ni^{3+}/O^-$ configurations and yielding a higher intrinsic TOF. In contrast, $H_2O$ incorporation partially compensates vacancy-induced $Ni^{3+}$ centers through local hydroxylation, forming Ni–O–H environments that facilitate faster electrochemical activation but reduce the intrinsic turnover per active site.

The TOF–Tafel behavior of the 88 % $O_2$ film aligns with reports that $Ni^{3+}/O^-$ ligand-hole configurations enhance lattice-oxygen-mediated OER pathways in Ni (oxy)hydroxides[90]. In contrast, hydroxylation reduces ligand-hole density and instead improves charge-transfer kinetics through Ni–O–H coordination, resulting in a lower Tafel slope despite a modestly reduced TOF and TOF×(Q/A).

## Conclusion

This work demonstrates that plasma-assisted deposition provides deterministic control over the defect landscape and electronic structure of NiO, enabling rational preconditioning of its transformation into the NiOOH OER-active phase. Oxygen-rich growth generates isolated and clustered Ni vacancies that strengthen Ni–O covalency and stabilize $Ni^{3+}$–$O^-$ ligand-hole states, thereby enhancing lattice-oxygen participation and accelerating OER-relevant electronic configurations. In contrast, controlled $H_2O$ incorporation selectively compensates vacancy-induced $Ni^{3+}$ centers, forming Ni–O–H environments that suppress deep divacancy-derived in-gap states while introducing shallow valence-band tails. These hydroxyl-driven modifications preserve the medium-range NiO lattice and promote rapid NiO → $Ni(OH)_2$ → NiOOH activation, improving charge-transfer kinetics.

Although DFT is used here to interpret experimentally accessible plasma conditions rather than to prescribe growth parameters a priori, the resulting electronic-structure framework defines the defect and electronic descriptors required for extension toward predictive plasma-guided design. Specifically, we identify two complementary defect-engineering pathways in NiO: vacancy-driven tuning of intrinsic electronic structure and hydroxylation-driven tuning of activation chemistry. By decoupling these effects during synthesis, plasma processing enables access to catalytic states that are difficult to achieve through post-synthetic approaches. More broadly, the integrated plasma–DFT–spectroscopy framework established here provides generalizable design principles for programming defect chemistry in transition-metal oxides, extendable to CoO, $LaNiO_3$, and layered or perovskite-type systems for developing robust, earth-abundant OER catalysts.

# Methodology

## Computational Details

First-principles calculations were performed within density functional theory (DFT) using the projector augmented-wave (PAW) basis set[25] as implemented in the Vienna *Ab initio* Simulation Package (VASP)[26]. Exchange correlation effects are described by the Perdew–Burke–Ernzerhof functional revised for solids (PBEsol)[92]. The PAW basis set treats Ni $3p^63d^84s^2$, O $2s^22p^4$, and H $1s^1$ as valence electrons. Because PBEsol alone cannot adequately capture strong electron-electron correlations–particularly in Mott insulators such as NiO–we treat the Ni $3d$ states with the DFT+$U$[27,28] method, applying an on-site Coulomb interaction $U$=4 eV.

All calculations are spin-polarized, and type-II antiferromagnetic (AFM) order is imposed for all systems. A plane-wave kinetic-energy cutoff of 520 eV is used, and Brillouin-zone sampling employs Monkhorst-Pack meshes of 8×8×8 for the 1×1×1 cell and 3×3×3 for the 3×3×3 supercell containing vacancies. Ionic relaxations proceed until the maximum residual force on any atom falls below 0.01 eV Å$^{-1}$.

## Sample Preparation

Ni foils were used as the substrates. They were polished using 15 μm and 8 μm silicon carbide (Starcke, Struers) and 0.5 μm alumina paste (MicroPolish Alumina, Buehler). The substrates were cleaned sequentially in an ultrasonic bath with acetone (Carl Roth), isopropanol (Carl Roth), and Millipore water for seven minutes each, dried using nitrogen, and loaded into the deposition chamber, which is a UHV chamber containing a magnetron sputtering source from MeiVac (MAK 2") connected in vacuum to an X-ray Photoelectron Spectroscopy System. The chamber is part of the DAISY-FUN cluster tool (Darmstadt's Integrated System for Fundamental Research).

Before each deposition, the chamber was evacuated to a base pressure of $2 \cdot 10^{-8}$ mbar. Ar and $O_2$ gas flows were introduced and varied in the plasma discharge, having a working pressure of $3 \cdot 10^{-2}$ mbar for the NiO thin film preparation. For h-NiO deposition, the water was enclosed in a Y-tube Pyrex glass, with a leak valve and needle valve to dose the water insertion into the deposition chamber. Water vapor was introduced in the 1.0 to $1.0 \cdot 10^{-4}$ mbar range, measured by a capacitive gauge sensor that measures independently of gas type and concentration. The partial pressures of $O_2$ and $H_2O$ can be estimated using equations 2 and 3, respectively. The working parameters used for NiO and h-NiO deposition for different thicknesses, $O_2$ concentrations, and $H_2O$ concentrations are summarized in Tables S1, S2, and S3.

$$O_2\% = \frac{f_{O_2}}{f_{O_2}+f_{Ar}} * 100 \qquad (2)$$

$$H_2O\% = \frac{P_{H_2O}-P_{base}}{P_{working}} * 100 \qquad (3)$$

Where $f_{O2}$ and $f_{Ar}$ are oxygen and argon gas flow rates, respectively, $P_{H2O}$ is water vapour pressure, $P_{base}$ is the base pressure, and $P_{working}$ is the working pressure.

For h-NiO thin films, the thickness is limited to 15 nm because the deposition rate is reduced considerably due to the reaction of $H_2O$ with the target erosion zone, leading to the formation of a less conductive oxide layer at the surface of the metallic nickel and a strong decrease of the target voltage associated with a reduction of the deposition rate. The same phenomenon is expected with higher oxygen concentrations, but in a reduced manner[38]. Too much water vapor ($>>5 \cdot 10^{-3}$) also saturates the turbomolecular pump, leading to pumping failure.

## Atomic Force Microscopy

The morphology and surface roughness of the NiO and h-NiO films were evaluated using a Bruker Dimension ICON AFM. The cantilever (type PPP-FMAuD-10, Nanosensors) was made of silicon and coated with gold on the detector side. It has a resonance frequency of 45-115 kHz, a force constant of 0.5-9.5 N/m$^2$, and a 10-15 µm tip height. During the analysis, the cantilever tip was brought near the center of the sample's surface. Using the Nanoscope Analysis 1.8 software, the tip was positioned in a region where the film was relatively flat and free from abrasions. The scan was done in a tapping mode using the optimum oscillation frequency while recording the image. The images were processed using NanoScope 9.3 software. The plane fit command was initially used to remove image distortions. It computes a single-order polynomial in the image and subtracts it from the image.[93] This was followed by the Flatten command, which removes tilt and low-frequency noise from the image.[94]

## Grazing-Incidence Wide-Angle X-ray Scattering (GIWAXS)

Grazing-Incidence Wide-Angle X-ray Scattering (GIWAXS) was carried out on a Rigaku Smartlab (Cu K$_\alpha$ radiation) diffractometer in a grazing incidence setup with $\theta_{source} = 0.3°$. Data were recorded between 5° and 90° (in units of 2θ) using a step size of 0.01°. This Analysis is limited to a 100 nm NiO thin film since a minimum film thickness is required to generate an observable diffraction signal. Hence, no data is available for the 15 nm h-NiO sample.

## Electrochemical Measurements

Electrochemical measurements were conducted in a three-electrode setup in a PECC-2 cell (Zahner-Elektrik GmbH & Co. KG) filled with 1 M KOH (Carl Roth). The NiO sample served as the working electrode, which was in contact with a gold-coated plate connected to the back of the cell. The sample was sealed using a rubber gasket with an internal diameter of 8 mm, corresponding to the film's geometric area exposed to the electrolyte. A Hg/HgO (1 M KOH) and a Pt wire were used as the reference and counter electrodes, respectively. Before any electrochemical measurements, the Hg/HgO reference electrode was calibrated versus a reversible hydrogen electrode (RHE, HydroFlex-Gaskatel).

The samples were exposed to 200 cyclic voltammograms (CV), cycles from 1.0 to 1.5 V (vs. RHE) with a scan rate of 100 mV/s to transform the NiO surface to Ni(OH)$_2$. After this activation treatment, 5 CVs were recorded from 1.0 to 1.9 V (vs. RHE) at a scan rate of 50 mV/s to verify the OER current density. After that, chronoamperometry (CA) measurement was done at 1.5 V (vs. RHE) for one hour to oxidize the Ni(OH)$_2$ into NiOOH. When the CA was finished, an additional 5 CVs from 1.0 to 1.9 V (vs. RHE) were recorded. The charge (Q) to calculate the n active sites was extracted from the oxidation wave of these scans.

## X-Ray Photoelectron Spectroscopy (XPS)

X-ray photoelectron spectroscopy (XPS) spectra were acquired using a SPECS PHOIBOS 150 spectrometer implemented at the DAISY-FUN cluster tool. It is equipped with an Al K$_\alpha$ X-ray source (monochromatic Focus 500 with XR50 M (SPECS), $hv$ = 1486.74 eV). Survey and detail spectra were measured in fixed analyzer transmission mode while choosing a pass energy of 20 eV (step size of 0.5 eV) for the survey and 10 eV (step size of 0.05 eV) for the core levels. The system was calibrated to 0.00 eV binding energy of the Fermi level of sputter-cleaned Au and Cu, as well as to the emission lines of metallic Au 4f$_{7/2}$ at 83.98 eV, Ag 3d$_{5/2}$ at 368.26 eV, and Cu 2p$_{3/2}$ at 932.67 eV binding energy with deviations $\leq 0.1$ eV. The data analysis was performed with CasaXPS, version 2.3.22.[95]. The core level spectra were fitted with a Shirley background and peaks of a GL(30) line shape. Intensity calculations were done based on relative sensitivity factors published by Scofield[96].

## Optical Emission Spectroscopy

The plasma was analyzed by optical emission spectroscopy using an OceanOptics spectrometer USB4000 in the UV-Vis range (200-1100nm). Once the desired gas flow (Ar, $O_2$, Ar/$O_2$, Ar/$O_2$/$H_2O$) was introduced into the plasma discharge, the emission spectra were recorded using OceanView 2.0 software and then processed and analyzed using OriginLab.

## Hard X-ray Absorption Spectroscopy–EXAFS Analysis and Fitting Procedure

EXAFS measurements were performed in transmission mode for the ex-situ sample characterization and in fluorescence mode for the operando characterization at the B18 beamline of the Diamond Light Source (Didcot, UK) during experiment SP38607–1, data set Ni_NiO, Ni_h-NiO, and Ni_NiO_insitu. Samples were analyzed at the Ni K-edge (8333.5 eV) (For technical specifications, verify the official B18 End station at the Diamond website[97]). A Si crystal cut along the (111) plane was used as the monochromator. Energy calibration was carried out using a metallic Ni foil as the reference standard. Pt-coated mirrors were used to suppress harmonic contributions[98].

EXAFS data were reduced using the Demeter package 0.9.26[99], and fits of the $k^2$-weighted data were performed in R-space using theoretical functions generated with the FEFF6 code[79]. To calculate the phases and amplitudes of the Feynman paths, a cubic rock-salt NiO CIF was employed[100]. For each sample, the first Ni-O and the second Ni-Ni shells were included. For each path, the initial fitting parameters were the mean-square disorder in the neighbor distances $\sigma^2$ and the interatomic distances r. A common value of energy shift $E_0$ relative to the theoretical value was included, and to account for occupancy effects, the amplitude reduction of $S_0^2$ was also refined. The data were Fourier transformed over k = 2.5 – 13 Å$^{-1}$ and back-transformed over R = 1 – 3.3 Å. A Hanning window was used in both transformations (default values: dk = 1.0 and dR = 0.0). Fitting was performed in both k- and R-space using the ranges specified in the Artemis reports, and employing k-weights of 1, 2, and 3 simultaneously. For both samples, the number of fitted parameters was kept below the number of independent points reported by Artemis for the selected k- and R-ranges, according to the Stern rule[101].

## Soft X-ray Absorption Spectroscopy

Ni L-edge and O K-edge spectra were measured in a high vacuum setup at the ISISS beamline of the German synchrotron facility BESSY II. All spectra were recorded in total electron yield (TEY) mode, which was collected from the analyzer nozzle of an XPS detector. The analyzer nozzle was closely aligned to the sample and biased to 90 V against it to increase the signal-to-noise ratio. The storage ring was operated in low alpha and decay mode with a maximum ring current of 8 mA. Measurements were performed on three different spots per sample, and for each sample position, two consecutive spectra per element were measured. The spectra were normalized to the mirror current. It has to be noted that the energy resolution at the Ni $L_{2,3}$- edge was low for the presented measurements due to the special operation mode of the storage ring. Therefore, the spectral shape differs from NiO spectra typically found in modern literature.

Nevertheless, when comparing the spectra taken on the samples to a reference spectrum of NiO taken at identical settings, qualitative changes can still be interpreted. As a reference sample, we have used NiO powder (Sigma Aldrich). Spectra of α/β-Ni(OH)$_2$ synthesized with a hydrothermal approach were taken from a previous beamtime.

### Transmission Electron Microscopy (TEM)

For the TEM measurements, the thin films were deposited on Cu grids. The TEM analysis was done using a JEOL JEM F2100 TEM equipped with an EDS detector (X-Max 80 SDD-Detektor, Oxford).

### Associated Content

Research data supporting Figures and Tables in the main text and supplementary information is made available via ZENODO.org under DOI: XXX.XXX.XXX (made available after acceptance).

### Author Contributions

**H.M.F.**: Writing – original draft, Coordination, Methodology, Conceptualization, Sample preparation, XPS collection, hard XAS collection, soft XAS analysis, Electrochemical Investigation and Analysis. **M.A.:** DFT simulations and Analysis, Writing. **C.Jr.M.**: XPS collection, AFM collection, XRD collection, investigation and Analysis, Writing. **A.T.:** hard-XAS data validation and Analysis, Writing. **G.W. and M.T.:** Soft-XAS collection, investigation, and Analysis, Writing. **E.A.:** HRTEM collection and Analysis. **T.A.K.:** hard-XAS collection. **C.C.:** XAS data validation and Analysis. **L.M.L.:** Funding acquisition. **J.P.H.**: Funding acquisition. All authors participate in the proofreading.

### Conflicts of interest

There are no conflicts to declare.

### Acknowledgments

**H.M.F**. **G.W**., **M.F.T**., and **J.P.H**. acknowledge the German Ministry of Research, Technology, and Space (BMFTR) for providing financial support within the H2Giga cluster project PrometH2eus (Fkz: 03HY105H and 03HY105E). **M**. **A**. acknowledges financial support from the Collaborative Research Center FLAIR (Fermi level engineering applied to oxide electroceramics), funded by the German Research Foundation (DFG) under Project-ID No. 46318420 −SFB 1548 (subproject A02). **E.A.**, and **L.M.L**. acknowledge the financial support of DFG in the framework of the Collaborative Research Center Transregio 270 (CRCTRR 270) project No. 405553726 (subproject Z01). **H.M.F.** thanks Diamond Light Source for allocating beamtime SP38607–1 at the beamline B18. **H.M.F. G.W.**, and **M.F.T.** acknowledge the allocation of synchrotron radiation beamtime at the ISISS beamline at the BESSY II electron storage ring. **G.W.** and

**M.F.T.** thank Axel Knop-Gericke and Detre Teschner for support during the beamtime. **M.A.** and **H.A.** thank Marcus Ekholm from the Theoretical Physics Division at Linköping University for his discussions on the calculations. **M.A.** thanks the Paderborn Supercomputing Center for allocating time for the calculations.

Graphical abstract

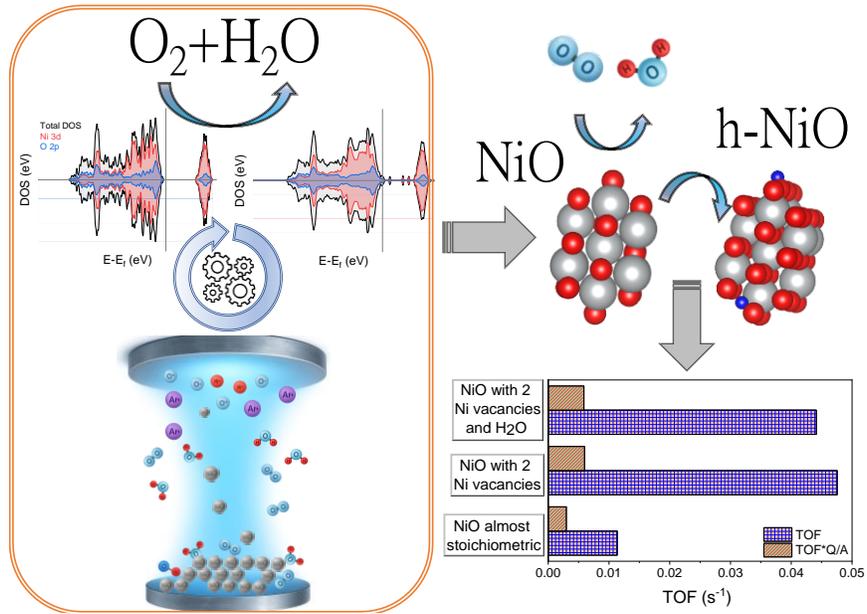

# Plasma-engineered Hydroxyl Defects in NiO: a DFT-Supported-Spectroscopic Analysis of Oxygen-Hole States and Implications for Water Oxidation

Harol Moreno Fernández[1*], Mohammad Amirabbasi[2*], Crizaldo Jr. Mempin[1,3], Andrea Trapletti[4], Garlef Wartner[5], Marc F. Tesh[5], Esmaeil Adabifiroozjaei[6], Thokozile A. Kathyola[7], Carlo Castellano[4], Leopoldo Molina-Luna[6], and Jan P. Hofmann[1*]

1) Surface Science Laboratory, Department of Materials and Geosciences, Technical University of Darmstadt, Peter-Grünberg-Straße 4, 64287 Darmstadt, Germany
2) Institute of Materials Science, Materials Modeling, Technical University of Darmstadt, Otto-Berndt-Straße 3, 64283 Darmstadt, Germany
3) Electrochemical Materials and Interfaces, Dutch Institute for Fundamental Energy Research, 5612 AJ Eindhoven, Netherlands
4) Dipartimento di Chimica, Università degli studi di Milano, Via Golgi 19, 20133 Milan, Italy
5) Max Planck Institute for Chemical Energy Conversion, Stiftstraße 34-36, 45470 Mülheim an der Ruhr, Germany
6) Advanced Electron Microscopy Division, Institute of Materials Science, Department of Materials and Geosciences, Technical University of Darmstadt, Peter-Grünberg-Straße 2, 64287 Darmstadt, Germany
7) Diamond Light Source, Didcot, Oxfordshire, OX11 0DE, United Kingdom

**Corresponding authors**

Harol Moreno Fernández – hmoreno@surface.tu-darmstadt.de
Mohammad Amirabbasi – amirabbasi@mm.tu-darmstadt.de
Jan P. Hofmann – hofmann@surface.tu-darmstadt.de

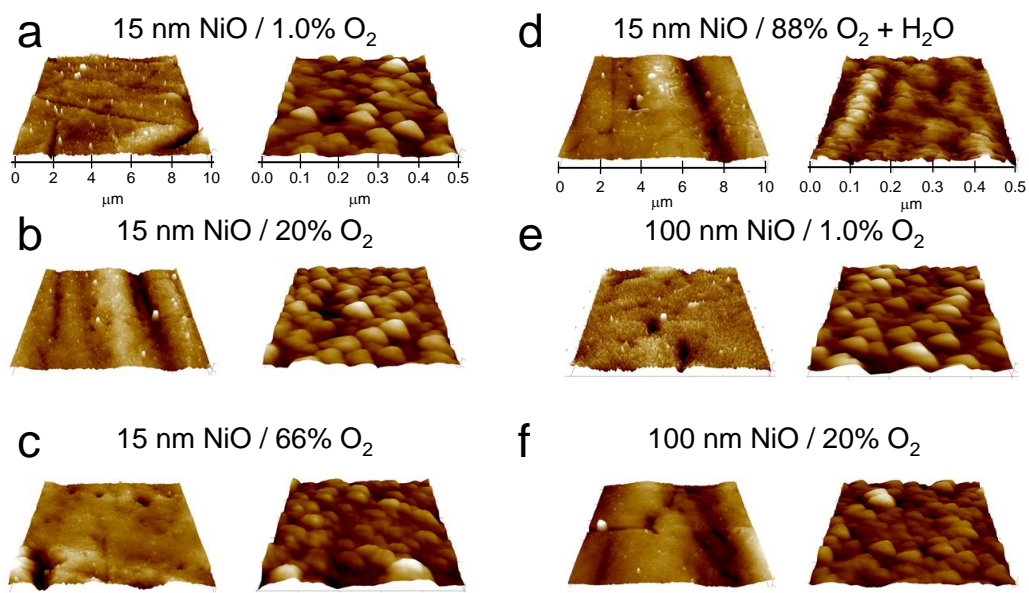

Figure S1. 3D AFM micrographs of NiO thin films of 15 and 100 nm and h-NiO thin films for different oxygen concentrations before any electrochemical conditioning (BC). The image was acquired using scan sizes of 10 × 10 µm² and 0.5 × 0.5 µm².

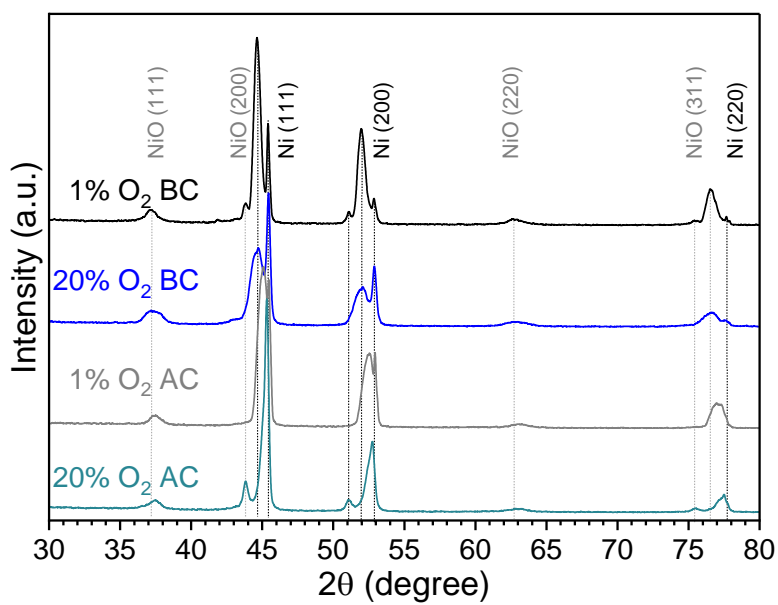

Figure S2. G-XRD diffractograms of NiO thin films of 100 nm prepared with 1.0% and 20% $O_2$ content before electrochemical conditioning (BC) and after 200 CV conditioning (AC).

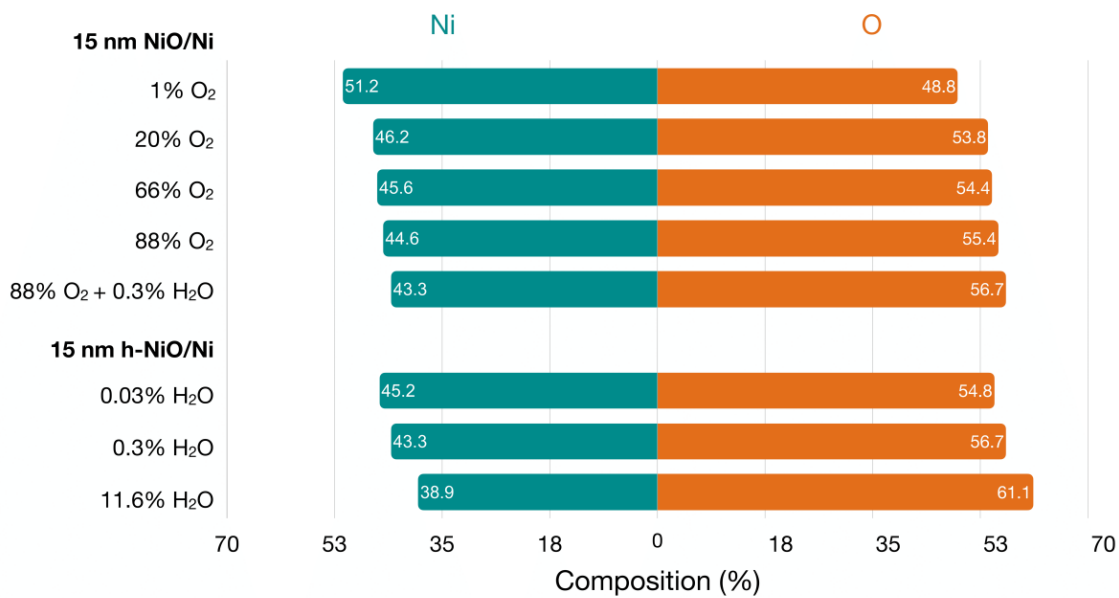

Figure S3. Atomic percentage composition derived from the XPS analysis of Ni 2p and O 1s core levels.

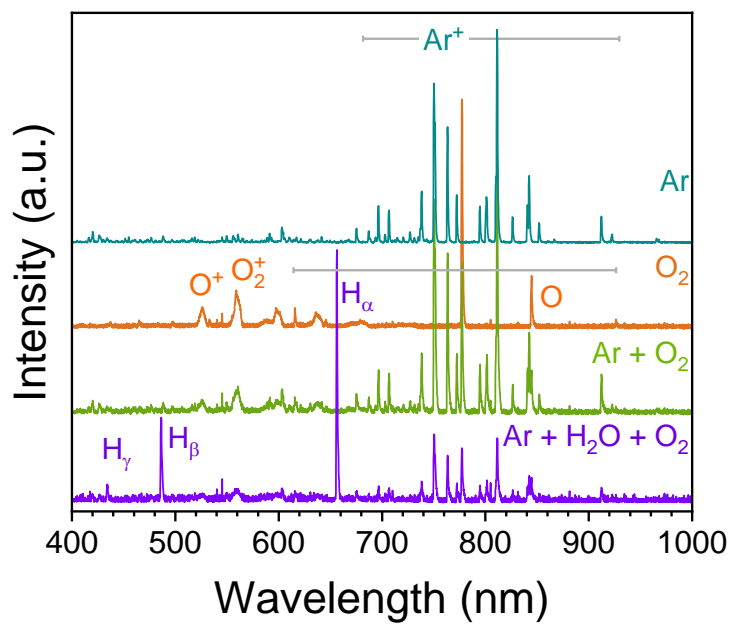

Figure S4. Optical emission spectra of the Ar, Ar+$O_2$, and Ar+$O_2$+$H_2O$ plasmas

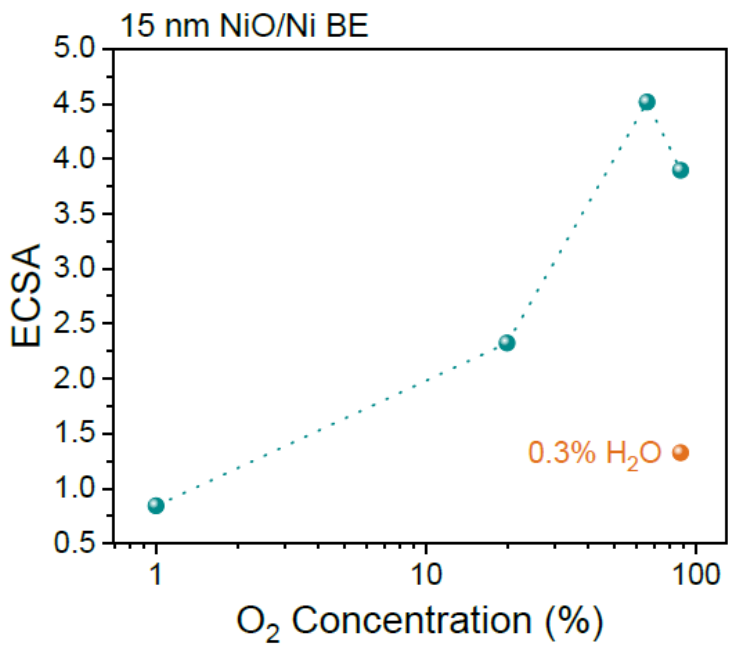

Figure S5. Electrochemical Active Surface Area (ECSA) obtained for thin films prepared with different oxygen concentrations and with 0.3% $H_2O$

Table S1 working parameters used to prepare NiO thin films of 15 nm and 100nm.

| Parameters | 1.0 % $O_2$ | | 20 % $O_2$ | |
|---|---|---|---|---|
| | 15 nm | 100 nm | 15 nm | 100 nm |
| Base pressure (mbar) | $1.0 \times 10^{-8}$ | | | |
| Working pressure (mbar) | $3.0 \times 10^{-2}$ | | | |
| Ar flow (sccm) | 19.8 | | 16 | |
| $O_2$ flow (sccm) | 0.2 | | 4 | |
| Deposition time (min) | 4 | 30 | 25 | 167 |
| Deposition rate (min) | 3.5 | | 0.6 | |
| Power (W) | 15 | | | |

Table S2 working parameters used for NiO thin films prepared with 1.0%, 20%, 88% $O_2$, and 88% + 0.3% $H_2O$.

| Parameters | 1.0 % $O_2$ | 20 % $O_2$ | 88 % $O_2$ | 88 % $O_2$ + 0.3% $H_2O$ |
|---|---|---|---|---|
| Base pressure (mbar) | $1.0 \times 10^{-8}$ | | | |
| Working pressure (mbar) | $3.0 \times 10^{-2}$ | | | |
| Ar flow (sccm) | 19.8 | 16 | 1.3 | |
| $O_2$ flow (sccm) | 0.2 | 4 | 10 | |
| $H_2O$ pressure (mbar) | - | | | $1.0 \times 10^{-4}$ |
| Deposition time (min) | 4 | 25 | 40 | 40 |
| Deposition rate (min) | 3.5 | 0.6 | 0.4 | |
| Power (W) | 15 | | | |

Table S3 working parameters used for NiO thin films prepared with 88% $O_2$ + different $H_2O$ concentrations.

| Parameters | 0.003 % $H_2O$ | 0.03 % $H_2O$ | 0.3 % $H_2O$ | 11.6 % $H_2O$ |
|---|---|---|---|---|
| Base pressure (mbar) | $1.0 \times 10^{-8}$ | | | |
| Working pressure (mbar) | $3.0 \times 10^{-2}$ | | | |
| Ar flow (sccm) | 19.8 | 2 | 1.3 | 2 |
| $O_2$ flow (sccm) | 0.2 | 4 | 10 | 4 |
| $H_2O$ pressure (mbar) | $1.0 \times 10^{-6}$ | $1.0 \times 10^{-5}$ | $1.0 \times 10^{-4}$ | $1.0 \times 10^{-3}$ |
| Deposition time (min) | 4 | 60 | 40 | 70 |
| Deposition rate (min) | 3.5 | 0.25 | 0.4 | 0.2 |
| Power (W) | 15 | | | |

*Plasma Diagnostics*

The optical emission spectra (OES) for the $Ar/O_2$ and $Ar/O_2/H_2O$ gas mixtures are shown in Figure S5, with peak assignments based on the NIST Atomic Spectra Database[1]. The discharge used for NiO deposition ($Ar/O_2$) is dominated by $Ar^+$ emission and oxygen-derived species assignable to $O^+$, $O^{2+}$, and $O^{3+}$, indicating an oxidizing plasma environment with a relatively high oxygen chemical potential. In contrast, the plasma used to grow h-NiO exhibits additional $H^+$ emission features, consistent with partial dissociation of water in the discharge. Although OES does not detect all fragments produced from $H_2O$ dissociation, the formation of ionic and radical species such as $OH^-$, $\dot{O}H$ and $\dot{H}$ in similar plasmas is well documented[2–6].

These reactive fragments interact with the surface during growth and are consistent with the incorporation of Ni–O–H coordination environments detected by XPS. The films remain polycrystalline under all plasma conditions, as confirmed by XRD/TEM. While EXAFS confirms that the medium-range NiO-like structure is preserved, it cannot resolve grain size or long-range ordering, and, owing to the fixed coordination numbers used during fitting, EXAFS is not sensitive to the exact concentration of Ni vacancies. XPS identifies mixed surface environments containing $Ni^{2+}$–O, $Ni^{3+}$–$O^-$, and Ni–O–H species, consistent with the defect-rich surface chemistry expected from plasma-grown NiO and h-NiO films.

The differences in plasma composition correlate with the defect chemistry extracted from DFT and spectroscopy. The more oxidizing $Ar/O_2$ discharge supports the formation of isolated and clustered Ni vacancies, producing oxygen-ligand-hole states and, at higher vacancy density, localized $Ni^{3+}$–$O^-$ configurations. Conversely, hydrogen-containing species in the $Ar/O_2/H_2O$ discharge contribute to surface hydroxylation, compensate vacancy-induced $Ni^{3+}$, and suppress the divacancy-related in-gap states predicted by DFT. These effects lead to a defect landscape consistent with XPS, VB, and EXAFS observations. They may contribute to the enhanced OER activity observed in h-NiO films, alongside improved electrochemical activation kinetics.